\documentclass[a4paper,fleqn,usenatbib]{mnras}

\usepackage[T1]{fontenc}
\usepackage{ae,aecompl}

\usepackage{graphicx}	
\usepackage{amsmath}	
\usepackage{subfig}
\usepackage{enumitem}
\usepackage{color}
\usepackage[dvipsnames]{xcolor}

\usepackage{newtxtext,newtxmath}
\usepackage{balance}
\usepackage{orcidlink}

\newcommand{\hone}{\textsc{Hi}}
\newcommand{\fft}{\mathcal{F}}

\title[$1/f$ noise in MeerKLASS]{Mitigating the effect of $1/f$ noise on the detection of the \hone \, intensity mapping power spectrum from single-dish measurements}

\author[M. O. Irfan et al.]{Melis O. Irfan\,\orcidlink{0000-0003-2021-7357}$^{1,2}$\thanks{E-mail: mirfan@myuwc.ac.za},
Yichao Li,$^{3}$
Mario G. Santos\,\orcidlink{0000-0003-3892-3073},$^{1,4}$
Philip Bull\,\orcidlink{0000-0001-5668-3101},$^{5,1}$
Junhua Gu\,\orcidlink{0000-0001-9765-6521},$^{6}$
\newauthor
Steven Cunnington\,\orcidlink{0000-0001-6594-107},$^{5}$
Keith Grainge,$^{5}$
Jingying Wang\orcidlink{0000-0002-5598-2668}$^{7,1}$
\\ \\
$^{1}$Department of Physics and Astronomy, University of Western Cape, Cape Town 7535, South Africa\\
$^{2}$Department of Physics and Astronomy, Queen Mary University of London, London, E1 4NS, UK\\
$^{3}$Department of Physics, College of Sciences, Northeastern University, Shenyang 110819, China\\
$^{4}$ South African Radio Observatory (SARAO), 2 Fir Street, Observatory, Cape Town, 7925, South Africa\\
$^{5}$Jodrell Bank Centre for Astrophysics, Department of Physics and Astronomy, The University of Manchester, Manchester M13 9PL, UK \\
$^{6}$National Astronomical Observatories, Chinese Academy of Sciences, 20A Datun Road, Beijing, 100101, China\\
$^{7}$Shanghai Astronomical Observatory, Chinese Academy of Sciences, 80 Nandan Road, Shanghai, 200030, China
}
\date{Accepted XXX. Received YYY; in original form ZZZ}

\pubyear{2022}

\begin{document}
\label{firstpage}
\pagerange{\pageref{firstpage}--\pageref{lastpage}}
\maketitle

\begin{abstract}
We present and compare several methods to mitigate time-correlated ($1/f$) noise within the \hone \, intensity mapping component of the MeerKAT Large Area Synoptic Survey (MeerKLASS). By simulating scan strategies, the \hone \, signal, foreground emissions, white and correlated noise, we assess the ability of various data processing pipelines to recover the power spectrum of \hone \, brightness temperature fluctuations. We use MeerKAT pilot data to assess the level of $1/f$ noise expected for the MeerKLASS survey and use these measurements to create realistic levels of time-correlated noise for our simulations. We find the time-correlated noise component within the pilot data to be between 10 and 20 times higher than the white noise level at the scale of $k = 0.04 \, {\rm{Mpc}}^{-1}$. Having determined that the MeerKAT $1/f$ noise is partially correlated across all the frequency channels, we employ Singular Value Decomposition (SVD) as a technique to remove both the $1/f$ noise and Galactic foregrounds but find that over-cleaning results in the removal of \hone \, power at large (angular and radial) scales; a power loss of 40 per cent is seen for a 3-mode SVD clean at the scale of $k = 0.04 \,{\rm{Mpc}}^{-1}$. We compare the impact of map-making using weighting by the full noise covariance (i.e. including a $1/f$ component), as opposed to just a simple unweighted binning, finding that including the time-correlated noise information reduces the excess power added by $1/f$ noise by up to 30 per cent.
\end{abstract}




\section{Introduction}
\label{sec:intro}

\hone \, intensity mapping, the measurement of unresolved, redshifted 21cm emission from neutral hydrogen, can be a powerful tracer of the formation of Large Scale Structure \citep{chang, loeb, peterson, chang10, masui, bull15}. Both interferometric and single-dish (auto-correlation) experiments have been developed for the purposes of tomographically mapping the 21cm brightness temperature fluctuations, with each method having its pros and cons \citep{battye13, santossSKA, bingo, fastLSS, chime}. Many of the challenges associated with measuring the \hone \, power spectrum are common to both interferometer and single-dish experiments, for example foreground removal or beam characterisation. There are important systematic effects that only affect either interferometers or auto-correlation experiments however. A particular example is correlated receiver gain fluctuations, which are an important contaminant for single-dish experiments, but can be reduced and controlled by interferometers. Such fluctuations can be seen as an effective multiplicative term, which is time-correlated over time scales longer than characterised by the `knee frequency' ($f_{k}$) and known as either $1/f$, pink or flicker noise. The receiver gain fluctuations ($\Delta G / G$) behind $1/f$ noise are a phenomenon found in a wide range of electronic systems, not only radio receivers, yet there is still a lack of understanding regarding their origin \citep{harper18, chen}. This $1/f$ noise can be modelled as an independent random Gaussian term with a power spectrum $S(f)$
for a single frequency channel ($\nu$) given by:
\begin{equation}
S(f) = \sigma_{{\rm{n}}}^{2} \left( 1 + \left(\frac{f_{k}}{f}\right)^{\alpha} \right),
\end{equation}
where $\sigma_{{\rm{n}}}$ is the thermal noise level, $f$ is the Fourier pair of the time axis of a stream of time-ordered data and the knee frequency ($f_{k}$) determines the frequency at which the $1/f$ slope (of gradient $\alpha$) meets the white noise level \citep{bsazy}. This noise term multiplies the sky temperature measured by the receiver.

However, with the onset of intensity mapping experiments intent on measuring a multitude of frequency channels with a single receiver, the possible correlation of the time-correlated noise across different frequency channels must be considered and parameterized  \citep{harper18}. In this work we build on this formalism by considering the impact of frequency-correlated $1/f$ noise on the \hone \, 3D power spectrum, with specific applications to single dish observations using the MeerKAT telescope \citep{2016mks..confE...1J}. The findings are crucial to guide a future large survey with MeerKAT: MeerKLASS \citep{mk17}, aiming to directly measure the anisotropic 21cm power spectrum at $\sim$ degree scales and above, over a wide redshift range. Moreover, MeerKLASS is a pre-cursor to the 21cm intensity mapping survey which will be conducted with the SKA-mid dishes and as such the findings in this work hold relevance for determining the optimum survey and data reduction strategies for 21cm intensity mapping with SKA-mid. A calibration pipeline was developed and applied to pilot observations from 2019 data \citep{wang21}, which have already been used to obtain a cross-correlation detection of the \hone \, power spectrum \citep{steve}. In this paper, we use observational measurements of receiver properties alongside actual scan strategies to produce realistic simulations of thermal and $1/f$ noise. These simulations help to determine the optimal configuration for the SKA-mid dishes in terms of scan strategies, time ordered data (TOD) reduction and map-making techniques to mitigate $1/f$ noise. 

The choice of scan strategy plays a pivotal role in the reduction of systematic noise. For $1/f$ noise the scan strategy (pattern and speed) determines the number of correlated data samples to be binned together into a single map pixel, whilst any changes in elevation during the scan effects the stability of the ground emission pick-up detected by the receiver. \citet{teg97oct} compares scan strategies in the context of the cosmic microwave background plus $1/f$ noise and determines that the optimal observational method is one which maximises the independent cross-linking between data samples. This can be done in practice by using a `fence' technique where regions close to each other on the sky are measured by the telescope across a time period shorter than the knee frequency, or by combining the data samples close to each other on the sky that have been measured by completely separate, and so independent, observation blocks. In this work we shall investigate two strategies; one with very little and one with moderate cross-linking. Whilst it is also possible to increase the number of data samples measured with uncorrelated noise properties by increasing the scan speed, we leave the simulation scan speed set to 5 arcmin per second. This value was chosen for the MeerKLASS pilot survey as it ensures that the pointing only changes by 10~arcmin within the 2~second integration time, which is sufficiently small compared with the primary beam size (approximately a degree at 1400 MHz) that it does not introduce significant smearing.

In terms of data reduction, several options for the mitigation of $1/f$ noise are available. As this time-correlated noise manifests as multiplicative gain drifts in the TOD, it can in principle be calibrated out. Some single dish receivers make use of noise diodes to inject a constant and stable signal into the TOD, which can be used as an internal calibrator \citep{gbtND, cbassND, hu21}. It is also possible to use a component separation technique, such as Singular Value Decomposition (SVD), on the TOD. Such methods can be used to decompose a matrix into components which are independent from each other. As the $1/f$ noise displays very different properties across frequency and time to the \hone \, signal, \citet{li21} have shown it is possible to use an SVD to help separate the two signals from each other. \citet{li21} use SVD cleaning on MeerKAT observational data of the South Celestial Pole and find that a 2-mode subtraction is enough to lower the knee frequency from around 0.1\,Hz to 0.003\,Hz. A final possibility, which we investigate in this work, is the level of $1/f$ mitigation that can be achieved from map-making using a weighting informed by the noise covariance matrix, as opposed to just a straightforward averaging \citep{teg97, sut10}.

As already mentioned, $1/f$ noise has previously been investigated in the context of intensity mapping simulations. In particular, \citet{bsazy} made an analysis considering a component completely correlated across frequency while \citet{harper18} considered noise partially correlated across frequency and its impact on the angular power spectrum in the context of an SKA wide survey. This work was further expanded in \citet{chen} to analyse the impact on cosmological parameters through Fisher forecasting, using an analytical model of the $1/f$ noise angular power spectrum. We build on these previous works and expand them with a detailed analysis of $1/f$ simulations in order to try to clarify the impact the $1/f$ noise will have on SKA-mid observations. In particular, we construct a realistic model of frequency-correlated noise using observational data from MeerKAT receivers taken during pilot surveys for intensity mapping. We then simulate the impact that both map-making techniques and proposed observational
scan strategies have on the measurement of the 3D \hone \ power spectrum (both in its cylindrical form and spherical average). Finally, we test how cleaning methods can alleviate the $1/f$ noise and check for possible signal loss using the cross correlation power spectra. In this context we present a simulation pipeline which will be useful for future works where realistic noise and scanning strategies are required.

The paper layout is as follows: in \autoref{sec:data} we outline the simulation pipeline and in \autoref{sec:postsvd} we assess how SVD cleaning the TOD impacts the magnitude of the $1/f$ noise within the 1D temporal power spectrum. The temporal power spectrum allows us to view the $1/f$ noise and white noise power levels as distinct from each other due to their different regions of dominance in $f$. \autoref{sec:method} details the map-making step and in \autoref{sec:res} we present the effect that SVD cleaning has on maps contaminated with $1/f$ noise through the use of power spectra. We conclude in \autoref{sec:conc}. Throughout the paper vectors are capitalised, and matrices are bold and in capitals.

\section{Simulation Data}
\label{sec:data}

\begin{table}
\centering
\begin{tabular}{||c c c||} 
 \hline
{\bf{Parameter}} & {\bf{HRD strategy}} &  {\bf{Random strategy}} \\
 \hline\hline
RA & $150^{\circ} - 175^{\circ}$ & $150^{\circ} - 175^{\circ}$ \\
Dec & $-1.0^{\circ} - + 8.0^{\circ}$ & $-1.0^{\circ} - + 8.0^{\circ}$ \\
Scan speed & 5 arcmin/s  & - \\
Slew speed & -  & 2$^{\circ}$/s \\
No. dishes & 60 & 60 \\
$T_{{\rm{sys}}}$ & 16\,K & 16\,K \\
$\delta t$ & 2\,s & 30\,s \\
Observation time & $16 \times 90$ min & $16 \times 90$ min\\
Frequency range & 971.2 -- 1171.2\,MHz & 971.2 -- 1171.2\,MHz \\ 
$\delta \nu$ & 0.2\,MHz & 0.2\,MHz \\
 \hline \hline
\end{tabular}
\caption{Pertinent simulation parameters for the Horizontal Raster Drift and Random Pointing survey scan strategies.}
\label{tab:sum}
\end{table}

The simulation data have been created to reflect the working set-up of the MeerKAT L-band (856--1711\,MHz) receivers. We simulated data from 60 dishes, as this is typically the number of working dishes available from the full array of 64 for an average observation block. As each dish has two receivers, measuring HH (horizontally polarised) and VV (vertically polarised), this gives 120 sets of simulated data. The frequency resolution was set to that of the L-band receivers (0.2\,MHz) and we chose to simulate 1000 data channels starting from 971.2\,MHz, as \citet{wang21} found significant satellite radio frequency interference under 970\,MHz. The L-band receivers have a frequency-dependent beam which, when approximated as a Gaussian primary beam, decreases in full width half maximum (FWHM) from 1.6$^{\circ}$ to 1.3$^{\circ}$ across the 971.2 to 1171.2\,MHz frequency range. Pertinent simulation parameters are summarised in \autoref{tab:sum}. The MeerKAT dishes can be set to follow numerous scan patterns; in this work we consider the Horizontal Raster Drift (HRD) and Random Pointing strategies. These scans will be divided into observing blocks of the same size in time, covering the same region sky. 

\subsection{Horizontal Raster Drift}

The HRD strategy is currently the survey mode being used to obtain MeerKAT data for \hone \, intensity mapping. It entails holding each dish at a constant elevation and scanning back and forth in azimuth, and is motivated by keeping the elevation-dependent power contributions from ground spill and atmospheric emission close to constant. In this strategy, scan lines from different blocks are made to cross so as to improve the overall relative calibration. A scan speed of 5 arcmin per second is used. Note that we do not consider here the impact of this relative (or redundant) calibration in the $1/f$ noise. Removing the $1/f$ noise through improvements in calibration through this crossing or noise diode injection will be left for future work. Moreover, as discussed later, the calibration in between observing blocks effectively removes the $1/f$ time correlations between blocks so that the lines crossing from different blocks should have little impact on improving map-making. This also means that we can effectively suppress the $1/f$ contribution by cross-correlating different blocks or dishes. In this instance, improvements in scanning strategies or map-making will not impact the final power spectrum on average. Instead, they will help to reduce the variance of the measurement.

The terrestrial contributions to the total system temperature are the receiver temperature, the atmospheric temperature and the ground emission pick-up. The receiver temperature is expected to change over frequency but to remain constant over long timescales with a magnitude between 6 and 8\,K \citep{wang21}. We use the receiver temperature models provided by the MeerKAT office, which are a function of dish, polarisation and frequency. The atmospheric contribution is the same for each receiver but changes across frequency and elevation. As the HRD scans at constant elevation though, the atmospheric contribution only changes across frequency and is constant in time. We calculate the simulated atmospheric contribution using typical pressure, humidity and temperature values for the MeerKAT site. The ground emission pick-up is different between the HH and VV polarisations and changes over frequency and elevation. As before, the constant elevation of the HRD scans means that we only need to use the MeerKAT office ground emission models in VV and HH across frequency. The combined elevation-dependant temperatures contribute between 4.4 and 5.5\,K, depending on the frequency channel, to the total emission temperature. The HRD strategy, therefore, has a combined terrestrial temperature of around 10-14\,K which varies across frequency and receiver but is well modelled as constant in time.

As the $1/f$ noise is a multiplicative factor, the magnitude of its fluctuations ($\delta_{1/f}$) scale with the total system temperature at each frequency:
\begin{equation}
T_{{\rm{sys,m}}}(\nu) = \delta_{1/f}  \left(T_{{\rm{rec}}} (\nu)+ T_{{\rm{el}}}(\nu) + T_{{\rm{CMB}}} + T_{{\rm{sky}}}(t, \nu) \right),
\end{equation}
where
\begin{equation}
T_{{\rm{sky}}}(t, \nu) =  T_{{\rm{HI}}}(t, \nu)  + T_{{\rm{Gal}}}(t, \nu)  + T_{{\rm{ExGal}}}(t, \nu),
\end{equation}
and $T_{{\rm{sys,m}}}$ is the measured system temperature. The only temperature contribution which changes in time for the HRD comes from the sky, namely the combined \hone \,, diffuse Galactic and extragalactic point source emission temperature contributions. When it comes to map-making for the HRD strategy, the temperature averages (across time) at each frequency channel are subtracted from the TOD to remove the terrestrial contributions to the measured system temperature. This is a common strategy in data analysis (removing a constant offset in time), which can also be accomplished through the SVD cleaning.

\subsection{Random Pointing}

For the Random Pointing strategy we first divide the area of sky we wish to observe into a hexagonal close packed grid of R.A and declination values with a spacing of 0.35$^{\circ}$. For the first 90 minute observation block we select a random (R.A, declination) pair from the grid and observe it for 30 seconds, then we select another random coordinate pair and slew to the next location. The MeerKAT dishes can slew at a speed of 2$^{\circ}$/s in azimuth/elevation so the slew time is calculated using the difference between one pointing and the next in the azimuth/elevation reference frame. If the difference in azimuth is larger than the difference in elevation then the azimuth difference is used to calculate the slew speed, while if the elevation difference is larger then that is what is used to calculate the slew speed. New pointings continue to be selected at random from the total grid until the 90 minutes observational time is used. The observed (R.A, declination) pairs are then removed from the grid options and the process is repeated again for the rest of the observational blocks. This random pointing is a closer representation of the strategy currently employed by the MeerKAT telescopes in interferometric mode, in terms of following a schedule of tracking observations. However, with just 30 seconds per pointing, the overheads in slew time will be quite large for MeerKAT, so this strategy is highly inefficient. Longer tracking times, would be more efficient but increase the 1/f noise dramatically.

   \begin{figure*}
    \centering
    \includegraphics[scale=0.55]{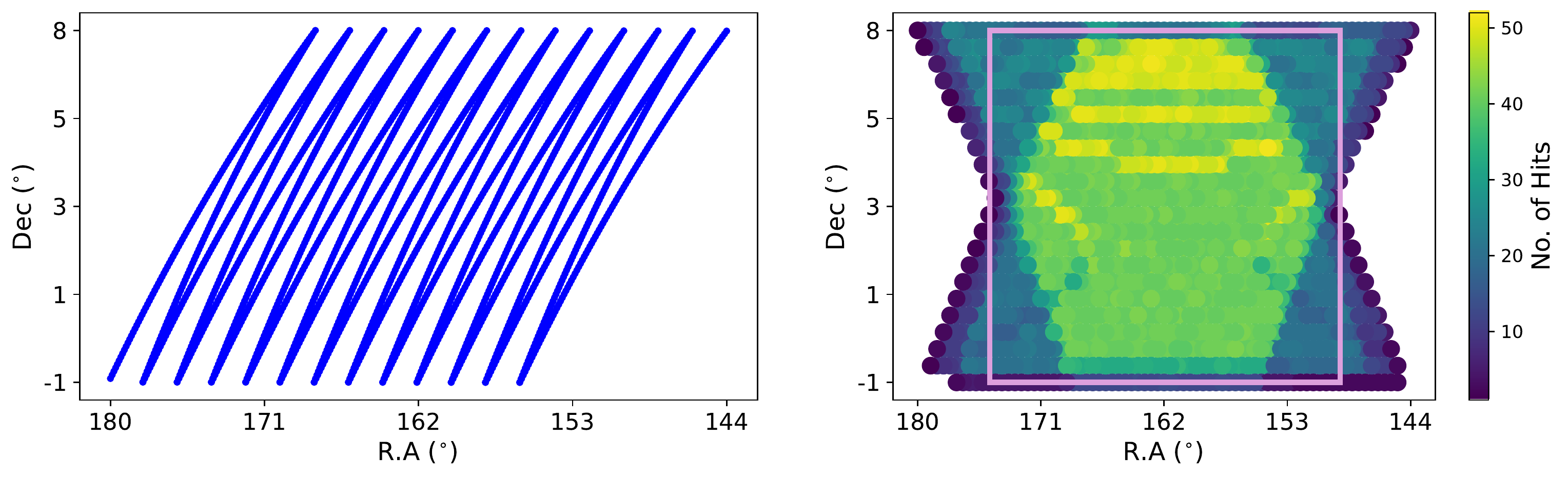}
    \includegraphics[scale=0.55]{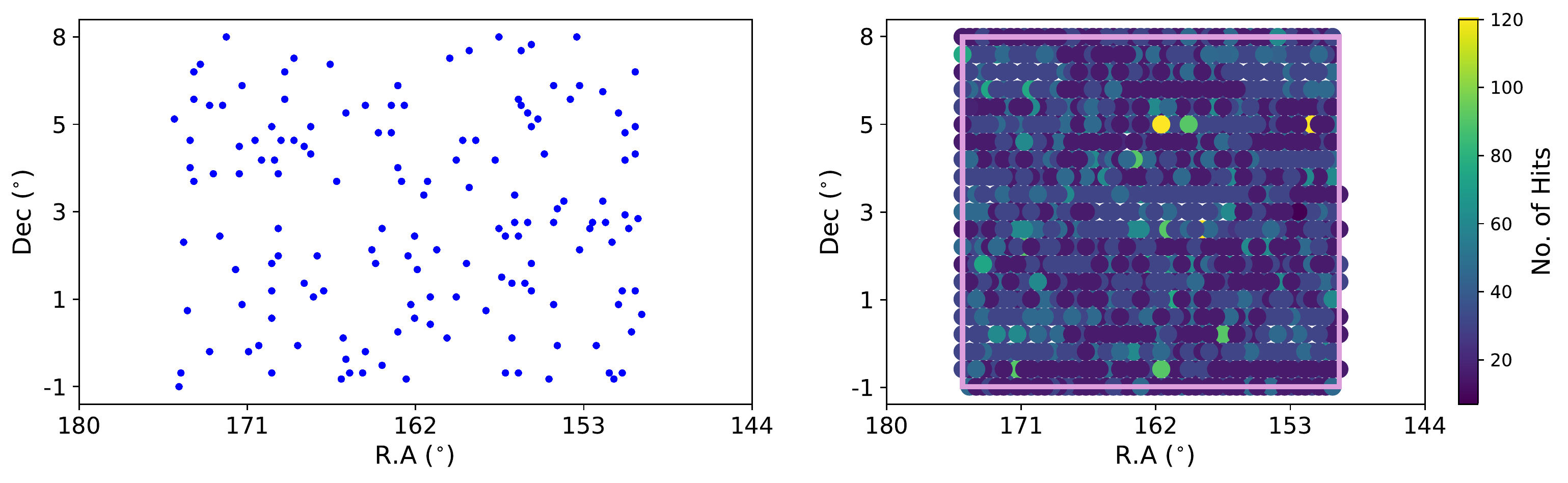}
    \caption{The region of the sky as covered by the HRD ({\it{top}}) and random ({\it{bottom}}) scan strategy for a single observation block ({\it{left}}) or for all 16 observation blocks ({\it{right}}). The purple rectangle denotes the region selected for use in this analysis.}
    \label{fig:area}
 \end{figure*}  

The simulated receiver temperatures used for the random strategy are identical to those used for the HRD. The elevation-dependant contributions, however, change over time for the random strategy, unlike for the HRD, as the dish elevation changes with each pointing. The pointing for each random observation block is used to provide the elevations necessary to calculate the atmospheric and ground spillover temperatures from the MeerKAT office models. These additional fluctuations, $\delta T_{{\rm{el}}}(t)$, can be up to 1\,K and so $\sigma_{1/f} \times \delta T_{{\rm{el}}}(t)$ will not be negligible and need to be included in the TOD. The MeerKLASS data reduction pipeline contains models for the elevation-dependant temperature contributions which we subtract from our empirical data \citep{wang21}. For these simulations we assume these models are perfect and so we completely subtract the time-varying, elevation-dependant temperatures at each frequency channel, after the TOD is generated but before map-making the random survey. The aim of this paper is to investigate the effect of $1/f$ noise on the \hone \, power spectrum, therefore this work will not deal with the impact of incorrect ground and atmospheric emission modelling.

A random pointing strategy would make using the interferometric visibilities a straightforward task. As it stands, the HRD scan strategy complicates this matter but not to the extent that it is no longer achievable; future papers will demonstrate the ability of MeerKLASS to measure simultaneously in auto- and cross-correlation mode. We do not consider more scanning strategies as these two sample what we believe is reasonable for a MeerKAT intensity mapping survey. A more complicated movement of dishes on the sky would improve the crossing between lines within the $1/f$ noise time scales but would severely affect the ground pick up due to strong elevation variations. On the other hand, an even more "classic" survey where we spend more time tracking each point, would accumulate $1/f$ noise on the scales of interest beyond any acceptable level. 

The simulated HRD/random pointing data consist of 16 observation blocks, each of which has a 90 minute duration. The 90 minute time duration was chosen as the temperatures of the MeerKAT LNAs have been shown to be stable over this time period \cite{wang21}. 16 observation blocks provides a high enough signal-to-noise ratio between the \hone \, signal and the instrumental Gaussian noise level.

\autoref{fig:area} shows the sky coverage for each strategy; the TOD samples are recorded every 2\,s for both scanning strategies but whilst the HRD strategy changes position on the sky with each sample, the random strategy observes each position on the sky for 30\,s (15 hits) before slewing to a new location. In order to assess the ability of SVD cleaning to reduce the impact of $1/f$ noise on the \hone \, cross-correlation power spectrum measurement we also needed to simulate the other frequency-correlated emissions expected to contribute to the total system temperature: Galactic and extragalactic foregrounds. We use the Global Sky Model software package \footnote{https://github.com/telegraphic/pygdsm} to simulate the foreground contributions at each frequency channel \citep{gsm}.

\subsection{Simulated 21cm signal}

For the 21cm signal we generate Gaussian realisations of a density field from an input power spectrum. A Gaussian overdensity field $\tilde{\delta}(\mathbf{k})$ is generated in Fourier space, populated with realisations of Gaussian random numbers with standard deviation $\sigma(k) = \sqrt{P(k)/2N_\text{vox}}$, 
where $N_\text{vox}$ is the number of voxels in the 3D field and a factor of two is required as the numbers are complex. $P(k)$ is the input power spectrum, which we obtain from \texttt{CAMB} \citep{Lewis} evolved to a redshift of $z=0.2$. Inverse Fourier transforming the field gives a periodic Gaussian realisation of the matter over-densities $\delta(\mathbf{x})$. To produce a \hone \, field from this we assume some linear bias $b_\text{HI}$ for the \hone \, and the mean brightness temperature $\overline{T}_\text{HI}$. The \hone \, temperature fluctuations are then given by 
\begin{equation}
    \delta T_\text{HI}(\mathbf{x},z) = \overline{T}_\text{HI}(z)b_\text{HI}(z)\delta(\mathbf{x},z)\,.
\end{equation}
For the bias we assume $b_\text{HI} = 1$ and we follow \cite{Switzer:2015ria} for the brightness temperature of the \hone, \, modelled by \citep{chang}
\begin{equation}
    \overline{T}_\text{HI}(z) = 0.39\,\text{mK} \, \frac{\Omega_\text{HI}(z)}{10^{-3}}\left[\frac{\Omega_\text{m}+\Omega_{\Lambda}(1+z)^{-3}}{0.29}\right]^{-1 / 2}\left[\frac{1+z}{2.5}\right]^{1 / 2}\,. \nonumber
\end{equation}
We assume $\Omega_\text{HI}= 6.2 \times 10^{-4}$. In this simplified model of \hone \, emission, we do not include the effect of redshift space distortions (RSD); leaving this for future work where we can fully investigate the effect that signal loss along the line of sight ($\nu$) has on measuring RSDs.  

The simulated \hone \, and foreground signals were convolved with an estimate for the MeerKAT beam. The convolution assumes a Gaussian beam where the beam FWHM changes across frequency as
\begin{equation}
\theta_{{\rm{FWHM}}}(\nu) = 1.2^{\circ} \times \frac{1280\,{\rm{MHz}}} {\nu},
\end{equation}
since the MeerKAT L-Band receivers are measured to have a FWHM of $1.2^{\circ}$ at 1280\,MHz \citep{kenda}.
Once more, no beam fluctuations were considered here in order to not complicate the understanding of the $1/f$ noise effects.

\subsection{Simulated noise}
White noise was generated for each receiver using a Gaussian distribution centred at zero. The white noise r.m.s level for each receiver is calculated using the receiver equation: 
\begin{equation}
\label{eq:wn}
\sigma_{{\rm{n}}} = \frac{T_{\rm{sys}}} { \sqrt{\delta t \, \delta \nu}},
\end{equation}
where $\delta t$ and $\delta \nu$ are the integration time and spectral resolution, respectively. Each dish has two linearly polarised receivers which are averaged together in order to form Stokes I at the map level. Therefore the white noise level in the maps can be calculated using \autoref{eq:wn} with an additional factor of $\sqrt{2}$ in the denominator.   

The $1/f$ noise was generated using the 2D $1/f$ simulator described in \citep{li21}. The $1/f$ noise in an individual receiver channel is correlated across the timestamps and simulated to have a temporal power spectrum which deviates from the constant power displayed by simple white noise in the form of a power law. This power law is characterised by the $\alpha$ parameter and the knee frequency ($f_{k}$), which is where the $1/f$ slope meets the white noise power level. The 1D Fourier transform of one TOD channel gives the noise power as a function of $f$
which is the reciprocal space of the time intervals. The $1/f$ noise across frequency channels may also be correlated \citep{harper18} i.e. the time-correlated noise is correlated over frequency as well. In this case the 2D Fourier transform is required to capture the full $1/f$ noise power information; the $1/f$ correlations in frequency are a function of $\tau$ which is proportional to 1/ frequency intervals between the different channels. We have chosen to model the spectral correlations using a power-law as well. The parameter $\beta$ quantifies how strongly correlated across frequency the $1/f$ noise is, ranging from not correlated at all ($\beta = 1$) to fully correlated across the entire frequency range ($\beta \to 0$). The 2D power spectra is calculated as:
\begin{equation}
{\bf{S}}(f, \tau) = \sigma_{n}^{2} \left( 1 + \frac{1}{K\delta \nu} \left(\frac{f_{k}}{f}\right)^{\alpha} \left(\frac{\tau_{0}}{\tau}\right)^{\frac{1-\beta}{\beta}} \right),
\end{equation}
where
\begin{equation}
K = \int d \tau \, {\rm{sinc}}^{2}(\pi \delta \nu \tau) \left(\frac{\tau_{0}}{\tau} \right)^{(1-\beta)/\beta},
\end{equation}
with $\tau_{0} = (N_{\nu} \delta\nu)^{-1}$,
and $N_{\nu}$ is the total number of frequency channels. \autoref{appendix:apA} details the measurement of the $\alpha, \beta$ and $f_{k}$ parameters used within this simulation.

In order to simulate the 1/f noise, we start by considering the 
2D power spectrum which is converted to Fourier coefficient amplitudes: 
\begin{equation}
X_{k} = \sqrt{\frac{S(f, \tau)}{\delta t \delta \nu}}.    
\end{equation}
These coefficients just contain amplitude information about the $1/f$ noise power; they have no phase information. The Fourier transform of real TOD produces a complex vector with both real and imaginary parts, therefore we need to add phase information to our amplitudes before performing the inverse Fourier transform. We add randomly generated phase information by multiplying the amplitudes by a unit amplitude Gaussian vector of complex numbers. Finally an inverse Fourier transform, enforcing that the resulting TOD should be real, can be used on the total (amplitude plus phase information) coefficients to produce the total noise timestream. These noise realizations are done independently for each receiver, dish and scanning block. While receivers and dishes should have statistically independent quantities, the $1/f$ noise could in principle be correlated across time scales much longer than the duration of each scan. However, in observations, the gains are fully calibrated at the beginning and end of each scan, which should calibrate out any large $1/f$ noise fluctuations and remove the very long time scale correlations.

\section{Quantifying the level of 1\,/\,\lowercase{\textit{f}} noise after SVD cleaning}
\label{sec:postsvd}
In this work we aim to investigate the effect that SVD cleaning has on the $1/f$ in both the TOD and the final maps. The basis of an SVD is that any real matrix ${\bf{A}}$ can be understood as the composite of two real orthogonal matrices ${\bf{U}}$ and ${\bf{V}}$:
\begin{equation}
{\bf{A}} = {\bf{U}} {\bf{\Lambda}} {\bf{V^{t}}},
\end{equation}
where ${\bf{A}}$ has dimensions $m \times n$, ${\bf{U}}$ has dimensions $m \times m$, ${\bf{\Lambda}}$ has dimensions $m \times n$ but only has values (the singular values of ${\bf{A}}$) on the diagonal and ${\bf{V}}$ has dimensions $n \times n$. For the TOD in a single observation block we have a two dimensional array of temperature samples at each frequency. The covariance matrix is calculated by taking the average over time of frequency-frequency pairs (or over pixels in the map). Then the Eigen-decomposition is performed on the frequency-frequency covariance matrix.

\begin{figure}
    \centering
    \includegraphics[scale=0.55]{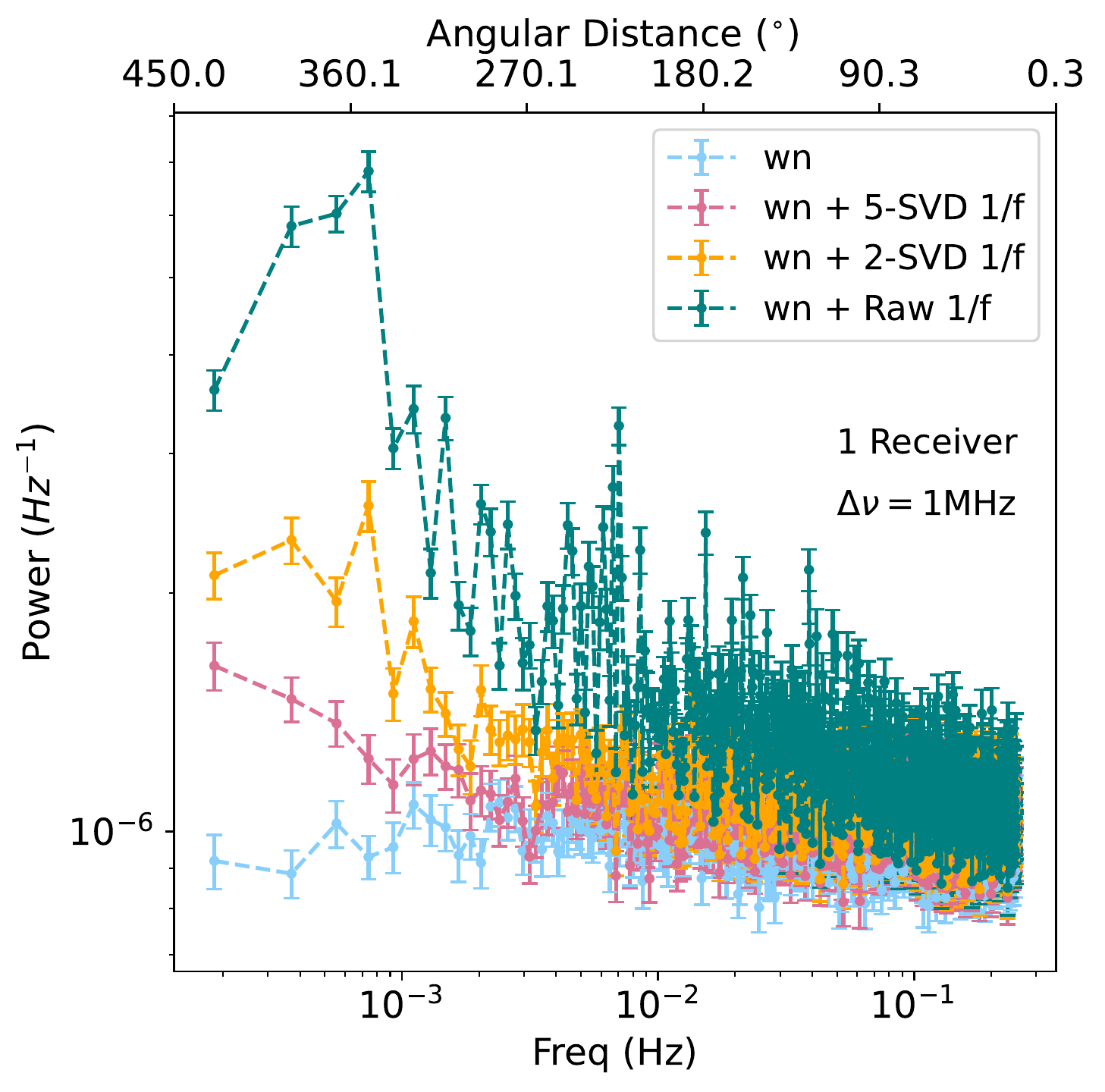}\\
    \caption{The temporal (1D) power spectrum of the HRD scan strategy TOD with $1/f$ noise at the raw and SVD cleaned levels for one observation block and one receiver. The white noise power level is also shown.}
    \label{fig:fftlow}
 \end{figure}

To quantify the effect of the differing $1/f$ noise levels (due to SVD cleaning) on the TOD we compare the Fourier transform of the various simulations. As, in this section, we are only interested in the relative level of the $1/f$ noise to the white noise level we compare TOD which only contain noise components, e.g. white noise and $1/f$ fluctuations only, no \hone \, emission and no foregrounds. In \autoref{fig:fftlow} we plot the 1D power spectra of the TOD after subtraction and division by the mean (see \autoref{eq:div}). If the TOD only contain white noise fluctuations, \citet{li21} demonstrated that the power for the normalised TOD will be a flat power spectrum at a level proportional to $1/ \delta \nu$. Although the simulation frequency resolution is 0.2\,MHz, to match the frequency resolution chosen for the MeerKLASS project, we chose to bin the simulated TOD channels into groups of 5 channels, giving a practical frequency resolution of 1\,MHz.

In \autoref{fig:fftlow} we show the temporal power spectrum for the HRD simulation. Binning our data to give a frequency resolution of 1\,MHz reduces the number of channels from 1000 to 200; the temporal power spectrum is calculated for each channel and then the mean power across the 200 channels is plotted, with the standard error on the mean giving the error bars. The blue curve shows the white noise power level at $10^{-6}$\,Hz$^{-1}$. Light pink and orange denote the $1/f$ power slope for the TOD that have been SVD cleaned by removing the first 5 principal modes and the first 2 principals modes, respectively. The green curve shows the raw level of $1/f$ noise expected in the TOD. It is clear that an SVD clean of the TOD decreases the $1/f$ noise power. We did not attempt to remove more than 5 SVD modes, despite the fact that the $1/f$ power is still larger than the white noise power level after a 5-mode clean, as we were concerned about possible over-cleaning i.e. removing the cosmological signal as well as correlated noise. The problem of over-cleaning is discussed further in \autoref{sec:res}.

The second x-axis shows the angular distance covered by the telescope as the scan progresses. If the scan strategy just drew a straight line on the sky this would also be the angular scale covered and so fluctuations at the Baryon Acoustic Oscillation (BAO) scale, around 1 deg, would see little impact from $1/f$ noise. However, as the HRD scan reverses direction after 18 degrees in azimuth, it re-visits points close together on the sky, so that this scale of angular distance is not equivalent to an actual angular scale on the sky. Noise measurements less than $\sim$ 20\,s apart can be seen to be independent. The typical $f_{k}$ value for the raw level of $1/f$ noise is 5\,mHz suggesting a typical timescale of 200\,s for $1/f$ noise dominating the total noise contribution. However, even if $1/f$ noise is not dominant its contribution is not less than 1 per cent of the white noise level until $\sim 1/10^{{\rm{th}}}$ of 200\,s, assuming an $\alpha$ value of 1, as the $1/f$ and white noise combine in quadrature. The choice of scan strategy has no impact on the power levels measured from time-ordered data meaning that both the HRD and random strategy produce almost identical power spectra and so we choose to only plot the HRD results in \autoref{fig:fftlow}.

\section{Map-Making}
\label{sec:method}

We now discuss the impact of map-making algorithms on the final $1/f$ noise properties of the power spectrum. The simulated TOD frequency channels are again first binned across 5 channels, thus increasing $\delta_{\nu}$ to 1\,MHz and reducing $N_{\rm{ch}}$ from 1000 to 200. We apply our map-making at each 1\,MHz frequency channel independently of the other channels. Even though our simulated $1/f$ noise is correlated in frequency, this correlated signal can be removed, or at least greatly reduced in magnitude, if a foreground component separation technique such as SVD is applied to the TOD. Therefore the main concern of the map-making is to mitigate biases and reduce errors in the \hone \,  auto- and cross-correlation power spectra caused by the time-correlated noise component in each individual frequency channel. Including frequency correlations in the noise covariance could improve the results but it would make the map-making much more computationally heavy (since then we could not do the map-making per frequency) and also it would introduce more inefficiencies due to the uncertainties in the full noise covariance matrix.

Following the approach of \citet{teg97, stompor, Hamilton}, we use a noise-informed map-making solution, assuming no prior knowledge of the \hone \, signal: 
\begin{equation}
 \label{eq:mmeq} 
X = ({\bf{A}}^{t} {\bf{C_{N}}}^{-1} {\bf{A}})^{-1} {\bf{A}}^{t} {\bf{C_{N}}}^{-1} Y, 
\end{equation}
where the vector of map pixels ($X$) is of length $N_{p}$, the vector of TOD ($Y$) is of length $N_{t}$, the pointing matrix (${\bf{A}}$) is of size $N_{t} \times N_{p}$ and the inverse noise covariance matrix, of dimensions $N_{t} \times N_{t}$, is used to weight the averaging of the TOD samples into map pixels.

In this paper, map-making is performed for all dishes and 16 observation blocks, so the length $N_{t}$ for the TOD samples includes the samples from each dish for all 16 of our observation blocks. 

\subsection{Simple binning}
\label{sec:bin}

In the case where there is no noise information or the data noise properties are believed to be purely Gaussian, \autoref{eq:mmeq} simplifies to just averaging the TOD samples into the appropriate map pixels:  
\begin{equation}
X = ({\bf{A}}^{t} {\bf{A}})^{-1} {\bf{A}}^{t} Y. 
\end{equation} 
In this case all that is required to make a map from the TOD is the pointing matrix. We use the Zenith Equal Area (ZEA) map projection making use of the \texttt{astropy} coordinate library \footnote{https://docs.astropy.org/en/stable/coordinates/index.html} selecting a pixel size of 0.458$^{\circ}$ (to match the pixel size in a HEALPix \footnote{https://healpix.sourceforge.io} map of $N_{\rm{side}}$ 128). The MeerKAT beam size at 971\,MHz is almost 1.6$^{\circ}$, giving between 2 and 3 map pixels per beam.

\subsection{Noise covariance}
\label{sec:mm}

What we denote as `map-making' in this work is the use of \autoref{eq:mmeq} explicitly including the noise covariance matrix. To estimate the noise vector we use an iterative process which starts from a naive map estimate: 
\begin{equation}
N_{0} = Y - {\bf{A}}({\bf{A}}^{t} {\bf{A}})^{-1} {\bf{A}}^{t} Y,     
\end{equation}
and is updated with each map estimate: 
\begin{equation}
N_{i} = Y - {\bf{A \, X}_{i}},     
\end{equation}
where the number of iterations used ($i$) is 2. In practice we have 16 observation blocks and 60 dishes. Each dish during a single observation block has noise properties independent from all other dishes and observation blocks. Therefore, the noise covariance is calculated per dish, per observation block and then all the temperature samples are simultaneously combined using this weighting and the full (for all observation blocks) pointing matrix.

Following \cite{sut09}, we use two approximations in order to implement \autoref{eq:mmeq} in the least computationally expensive manner. Firstly, in place of calculating a covariance matrix we use the circulant matrix approximation. If the noise covariance matrix is circulant, i.e. each row has identical elements to the row above, just rotated one element to the right, then:
\begin{equation}
\label{eq:noisepow}
{\bf{C_{N}}}^{-1} Y = \fft^{-1}(\fft(Y) / \fft({\bf{C_{N}^{1n}}})),
\end{equation}
where ${C_{N}^{1\rm{n}}}$ is the first row of the noise covariance matrix, $\fft{}$ denotes the Fourier transform and $\fft({C_{N}^{1\rm{n}}})$ is the statistical average over all realisations of the noise power spectrum (note that matrix division here corresponds to an element by element division). In practice only one measurement of the noise vector exists per dish and so we fit the $1/f$ noise model to the single noise power spectrum for each dish \citep{Dupac02}. The $1/f$ noise model in 1D is simply: $c_{t} (1 + (f_{k}/f)^{\alpha})$ and we leave all three parameters free to be fit. This approximation relies on the fact that the noise covariance is a circulant matrix. Although the noise covariance matrix for $1/f$ noise is not strictly circulant, \cite{teg97oct} detail how it can be decomposed into the sum of two matrices, one circulant and one sparse. \cite{teg97oct} explain how the circulant matrix dominates the noise covariance matrix operations as long as the rank of the noise covariance matrix is far larger than the sample difference over which the correlations fall to zero. For these simulations we are using 16 observation blocks of 90 minutes (where the sample rate is every 2 seconds) and 120 receivers; giving a noise matrix covariance rank of the order of $10^{6}$. The $1/f$ noise is correlated for a single receiver and only within the 90 minute observation block, making the noise correlations only hold over a sample difference of order $10^{3}$.

Therefore, as our noise covariance matrix can be decomposed into two matrices, one of which is sparse and the other of which is dominant and circulant, we can proceed by using the noise power spectrum approximation from \autoref{eq:noisepow}. Thus making the implemented map calculation differ from Eq.~\ref{eq:mmeq} like so: 
\begin{equation}
\label{eq:act}
X = {\bf{A}}^{t} {\rm{B}} \left( {\bf{A}}^{t} \rm{\bf{D}} \right)^{-1},
\end{equation} 
where ${\rm{B}} = \fft(Y) / \fft(\tilde{N})$ and ${{\rm{\bf{D}}}} = \fft( {\bf{A}}) / \fft(\tilde{N})$, where each row of the matrix  ${\rm{\bf{D}}}$ is constructed one at a time. The second approximation is used in place of matrix inversion; from \autoref{eq:act} we can see that the term in the brackets requires inversion. However in place of explicit matrix inversion we use the iterative generalized minimal residual function \texttt{lmgres} in \texttt{python}.

For experiments which aim to measure large regions of sky at high resolution and possibly in more than one Stokes parameter as well, the maximum likelihood method of map-making used in this paper might be prohibitively slow to implement. An alternative method of map-making, which is still capable of obtaining the optimum solution but at a far lower computational expense, called \texttt{DESCART} \citep{sut09,sut10} has been proposed - specifically in the context of mitigating the effects of $1/f$ noise on the measurement of the cosmic microwave background (CMB) angular temperature and polarisation power spectrum. The typical frequency for experiments which target a measurement of the CMB anisotropies is around 100\,GHz and at this frequency the CMB is dominant compared
to foregrounds, making systematics a more important consideration. \texttt{DESCART} divides the TOD into smaller time periods over which the $1/f$ noise can be approximated as an additive offset, which can then be calculated and subtracted. \cite{sut09} demonstrate a cut of computation time between 5 and 22 times when comparing \texttt{DESCART} to the standard maximum-likelihood map-making method. However, for intensity mapping, we only observe in Stokes I and at low resolution, therefore we choose to implement the full maximum likelihood solution.

\section{Results}
 \label{sec:res}

In this section both spherically and cylindrically averaged auto- and cross-correlation power spectra are used to understand the impact of the various levels of $1/f$ noise on the detection of the \hone \, signal. Spherically averaged power spectra show power as a function of $k$ where $k = \sqrt{k_{x}^{2} + k_{y}^{2} + k_{z}^{2}}$ while cylindrically averaged power spectra show the distribution of the power both perpendicular ($k_\perp = \sqrt{k_{x}^{2} + k_{y}^{2}}$) and parallel ($k_\parallel = k_{z}$) to the line of sight. 

     \begin{figure}
    \centering
        \includegraphics[scale=0.5]{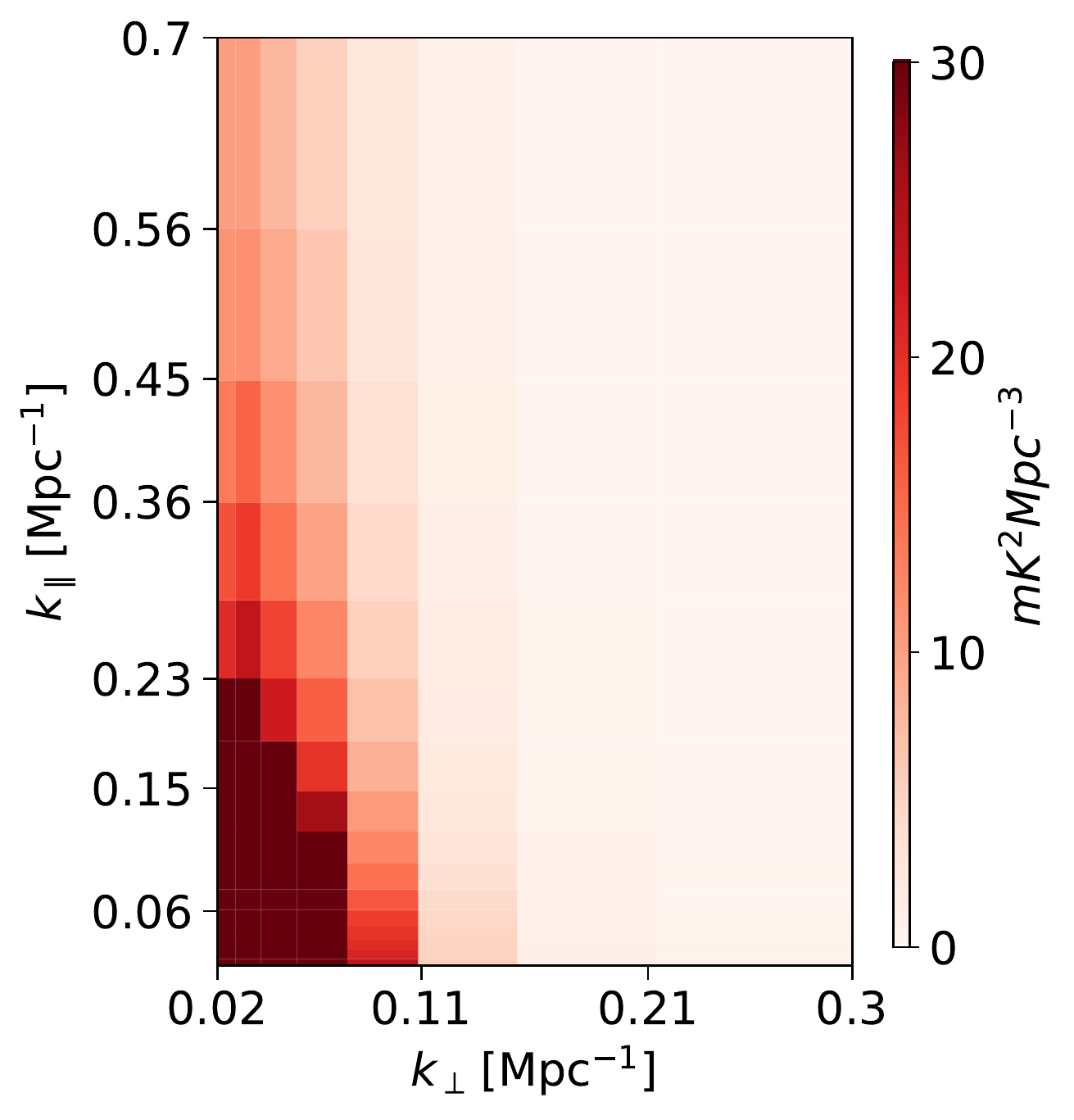}
    \caption{Cylindrically-averaged auto-correlation power spectrum for the \hone \, signal only. We impose a maximum limit of $30\,\text{mK}^{2}{\rm{Mpc}}^{3}$ to avoid saturation from the dominant scales.}
    \label{fig:ps2dh1}
 \end{figure}

      \begin{figure}
    \centering
        \includegraphics[scale=0.7]{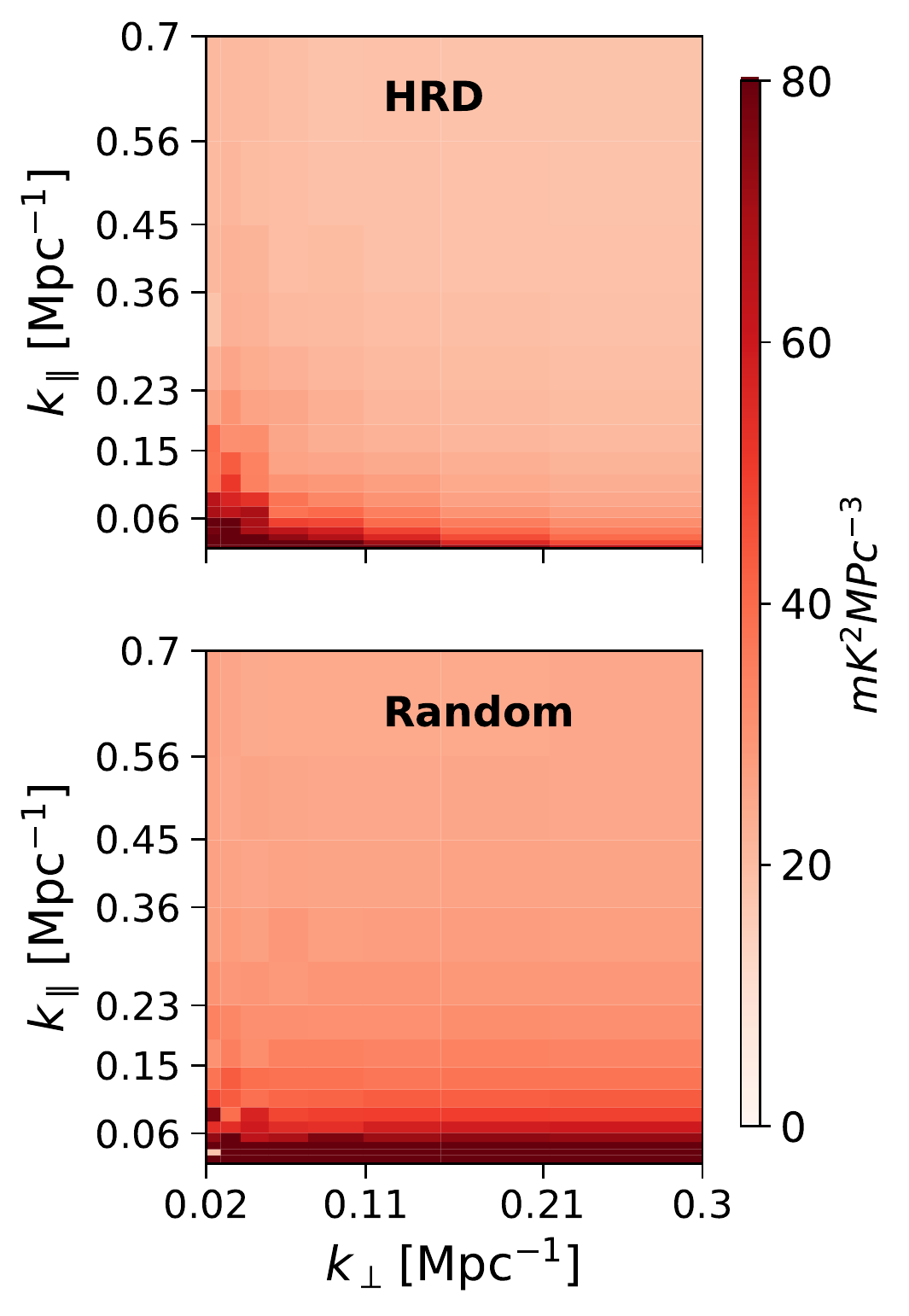}
    \caption{Cylindrically-averaged auto-correlation power spectrum for the white noise plus $1/f$ signal only. We impose a maximum limit of $80\,\text{mK}^{2}{\rm{Mpc}}^{3}$ to avoid saturation from the dominant scales.}
    \label{fig:ps2wn}
 \end{figure}

\begin{figure}
    \centering
        \includegraphics[scale=0.7]{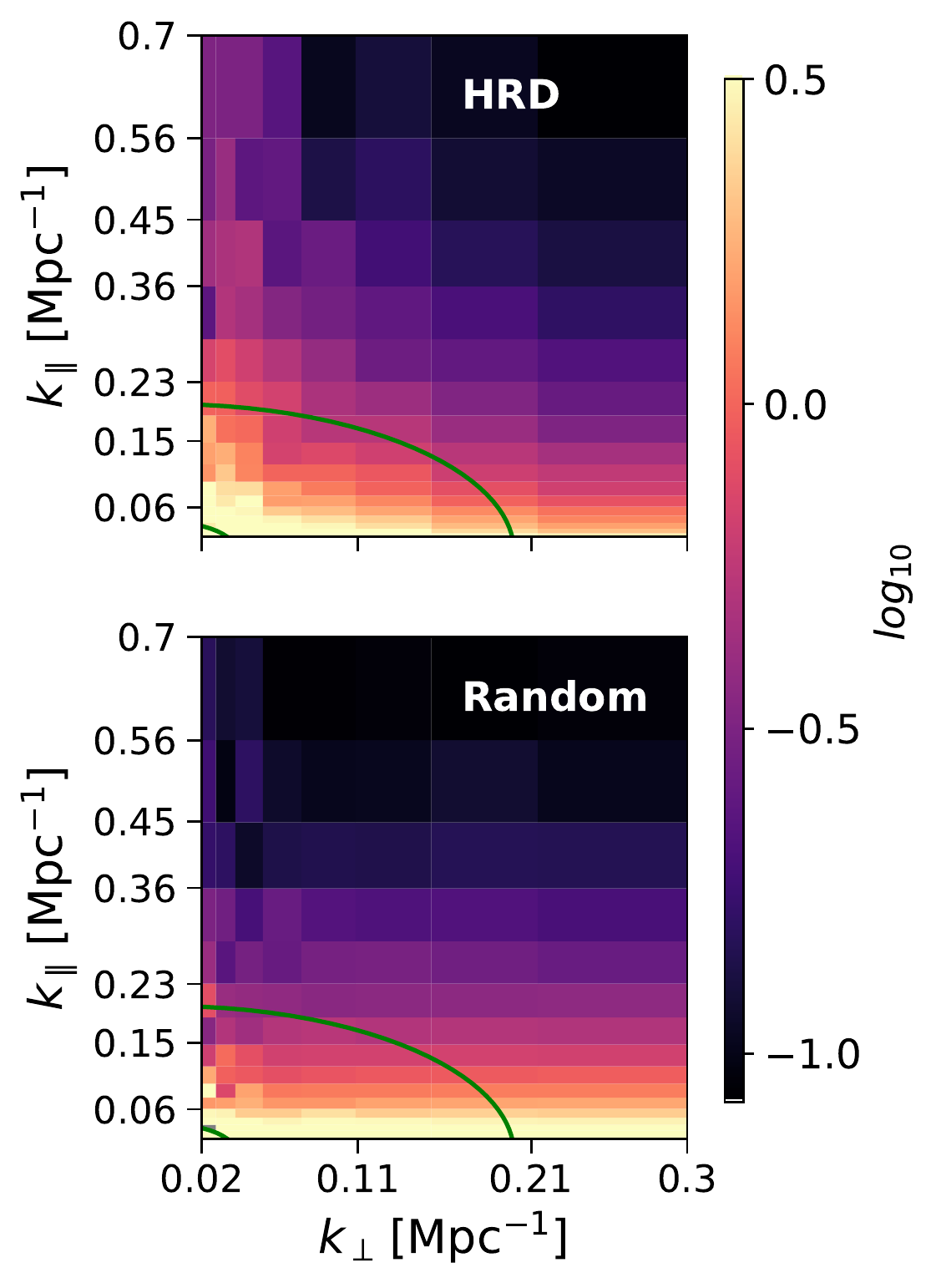}
    \caption{Cylindrically-averaged auto-correlation power spectra of $1/f$ noise power divided by the white noise power. {\it{Top}}: the HRD scan strategy, {\it{bottom}}: the random scan strategy. We use a logarithmic scale for the colour bars, imposing an upper limit of $\rm{log}_{10}(0.5)$ to highlight regions of $1/f$ dominance, and denote the area of $k$-space containing the three largest BAO peaks using two solid green contour lines.}
    \label{fig:ps2d}
 \end{figure}

For each scenario, e.g. TOD containing \hone, \,white noise and the raw $1/f$ noise, we create 20 realisations in order to plot the mean of the power spectra results. The functions used to calculate the power spectrum from the temperature cubes (RA, declination, frequency) have been taken from \texttt{FastBox} \footnote{https://github.com/philbull/FastBox}. In \autoref{fig:ps2dh1} we present the cylindrically averaged power spectrum for the \hone \, signal only, which shows large-scale frequency and angular correlations as the power is largest at small $k{_\perp}$ and $k{_\parallel}$ values. Both \autoref{fig:ps2wn} and \autoref{fig:ps2d} show the effect $1/f$ noise has on the cylindrically averaged power spectra. \autoref{fig:ps2wn} shows the white noise plus $1/f$ noise level for both the HRD and random survey while in \autoref{fig:ps2d}, the top panel is the $1/f$ noise power divided by the white noise power for the HRD scan strategy and the bottom panel image is the same but for the random scan strategy. For the random strategy we remove the perfect time-varying model for the elevation-dependant temperature before using the data. The $1/f$ power level is shown after division by the white noise level, in order to demonstrate the impact of solely the time-correlated noise component on the overall error budget. On both plots we highlight the region in $k$-space containing the three largest BAO peaks using green contour lines. 

The 2D power spectrum for the random scan strategy will always have a fractionally higher white noise power level than the equivalent HRD power spectrum as observation time is lost due to telescope slewing; on average around 15 per cent of the total observation time is lost. For these simulations the HRD white noise power level sits around 15\,${\rm{mK^{2}Mpc^{-3}}}$, while the random survey has a white noise power of $\sim 20\,{\rm{mK^{2}Mpc^{-3}}}$. A lower white noise level is not a motivator for selecting the scan strategy however, as the total observational time can always be increased. For the HRD strategy significant time is dedicated to observing the sky beyond the central rectangular region investigated in this work; in practice this time is not wasted as it can be combined with observations from an adjacent field. 

The random pointing strategy means that data points spatially close together could have any possible separation in time between 30\,s and 90 minutes and so the $1/f$ noise is not constrained to lie within large angular scales/small values of $k{_\perp}$. Whereas for the HRD strategy small values of $k{_\perp}$ mean longer time scales and so this region of the 2D power spectra is where we see the additional power of the $1/f$ noise. Both the top and bottom plots show the $1/f$ noise increasing the power at low values of $k_{\parallel}$, due to the frequency correlated nature of the $1/f$ noise. The random scan strategy results in a spreading of the $1/f$ noise throughout $k$-space, whereas the HRD strategy contains the $1/f$ noise to within the largest modes. At the largest angular scales ($ k _{\perp}\sim 0.02$) the random scan strategy shows a slightly smaller ratio of $1/f$ noise over the white noise level than the HRD scan, however this is not the case for the majority of the region in $k$-space containing the three largest BAO peaks.

 \begin{figure}
    \centering
    \includegraphics[scale=0.6]{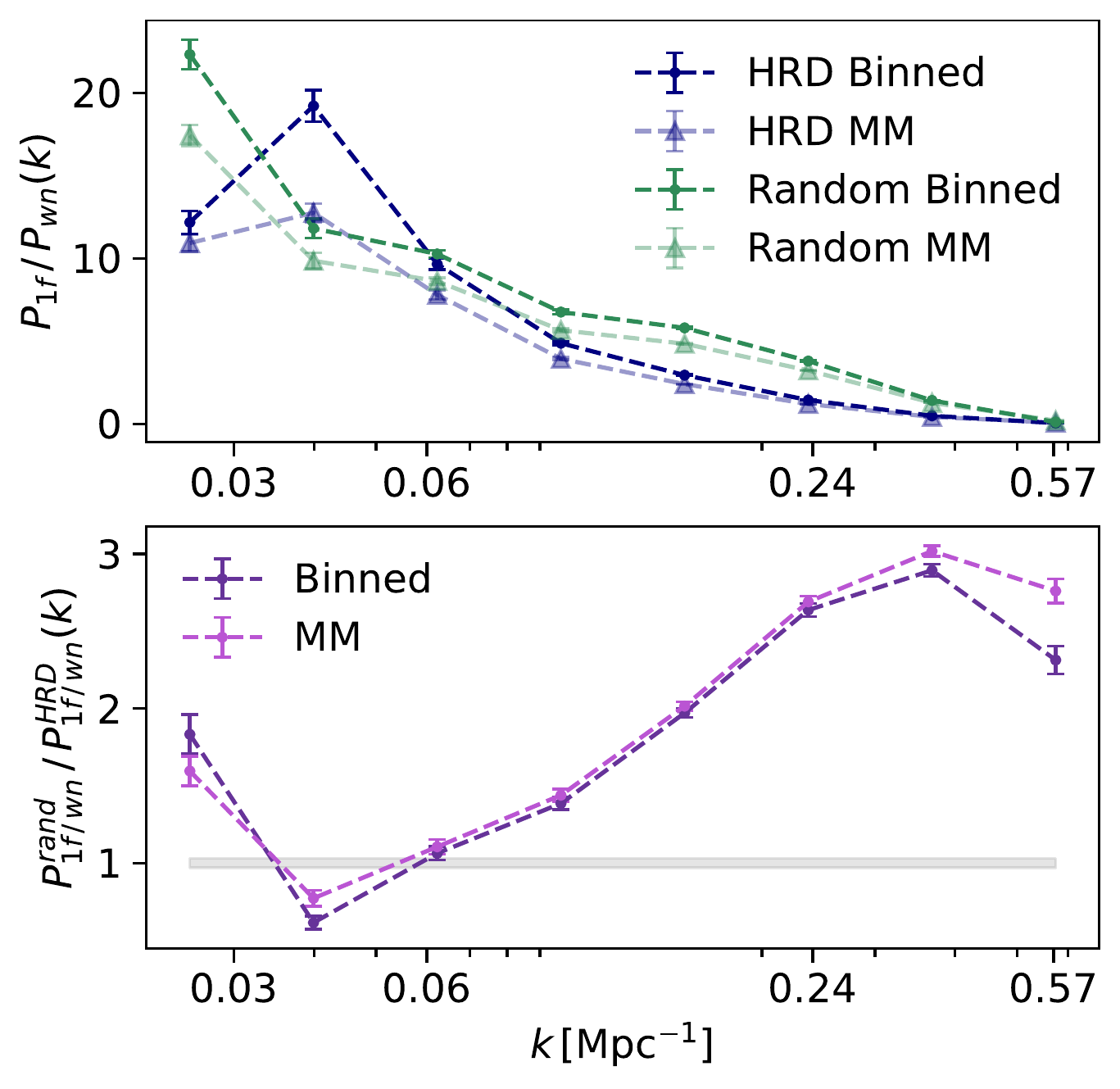}
    \caption{{\it{Top:}} Auto-correlation spherically-averaged power spectra for $1/f$ and white noise ratios, both scan strategies as well as both binned and noise-weighted maps are considered. {\it{Bottom:}} A comparison between the $1/f$ and white noise ratios for both the HRD and random survey strategy. Scales larger then 0.03 Mpc$^{-1}$ are not shown due to the increased noise on the power measurements arising from the compact size of the survey regions.}
    \label{fig:psraw}
 \end{figure}

In \autoref{fig:psraw} the spherically averaged 1D auto-correlation power spectra are plotted for several scenarios; in place of the true power level we choose to plot the ratio between the $1/f$ noise power spectrum for each scenario and the power spectrum of the white noise only. Each point on the power spectra represents the average power for 20 simulations and the error bars are the standard error on that mean. \autoref{fig:psraw} reveals that the raw level of $1/f$ noise found in MeerKAT receivers would raise the auto-correlation noise power spectrum measurement to around 20 times (depending on the scan strategy used) the white noise level. Map-making can be seen to provide on a average around a 30 per cent decrease in excess power due to $1/f$ noise, at the worst effected $k$-scales for both strategies; reinforcing the idea that noise-weighted map-making is preferable over a simple averaging/binning of the TOD into maps. The bottom panel of \autoref{fig:psraw} shows the ratio between the $1/f$ and white noise fraction for the random and HRD strategy. The power = 1 line is highlighted in gray to draw attention to the threshold below which the random survey outperforms the HRD survey. It is clear from this panel that using a random scanning strategy would result in a higher level of $1/f$ noise systematic noise above the Gaussian noise level for the majority of the region of $k$-space containing the largest three BAO peaks. Although the random strategy outperforms the HRD specifically at $k=0.04\,{\rm{Mpc}}^{-1}$, this advantage is not only fractional but also not seen for any other scale and so from this point onward we shall only consider the HRD survey results.

  \begin{figure}
    \centering
    \includegraphics[scale=0.6]{ 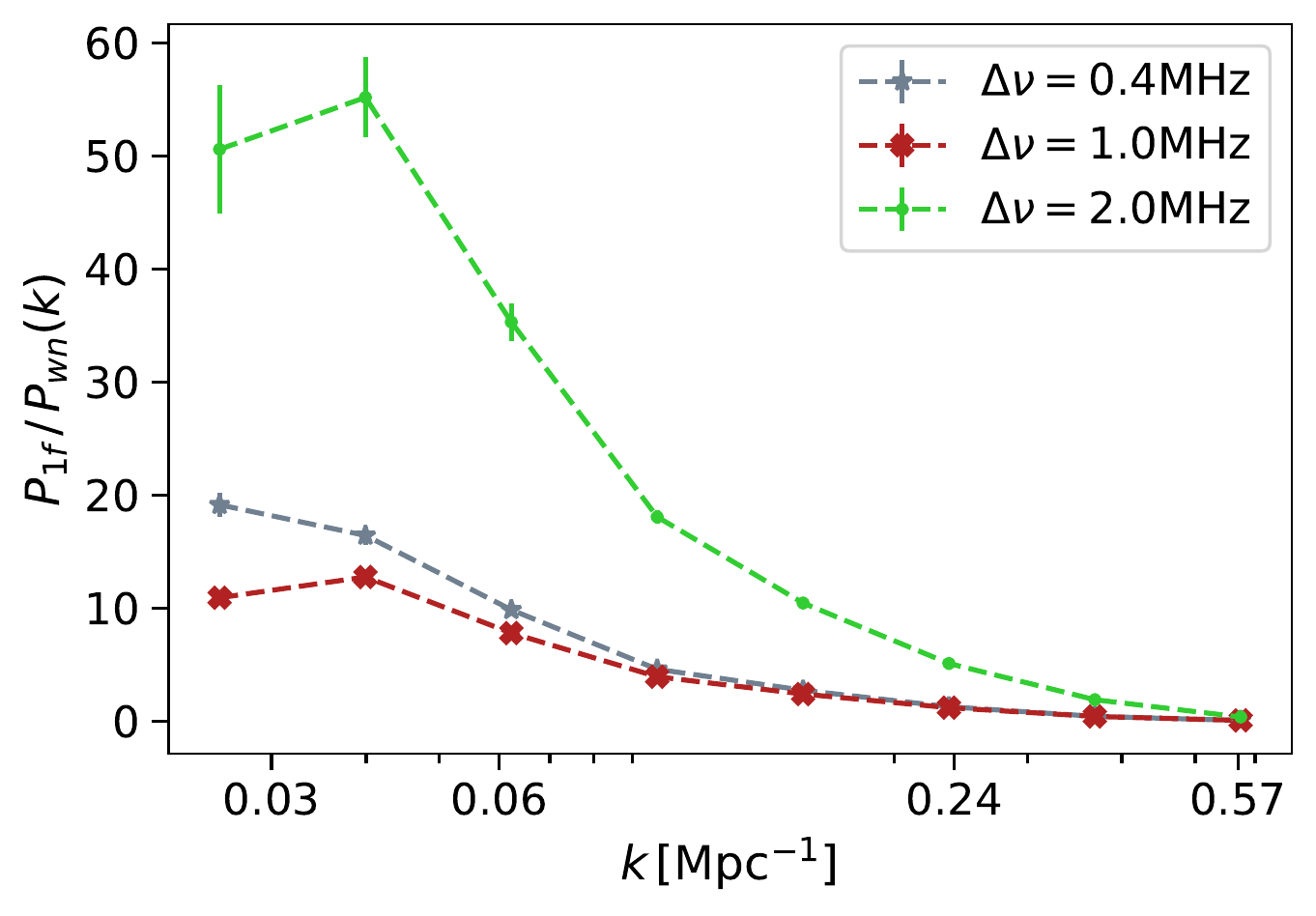} \\ 
    \caption{Power spectrum ratio between $1/f$ noise and the white noise maps for the HRD scan data for different frequency bins. The TOD were made into 2D maps at each frequency using the noise covariance matrix method of map-making, as opposed to simple binning.}
    \label{fig:psRatio}
 \end{figure}

We also consider how to bin the frequency channels. As discussed in \autoref{sec:method}, the frequency channels have been averaged over 5 channels, thus reducing the number of channels from 1000 to 200. This frequency averaging reduces the white noise level and so the knee frequency in each channel map and it also reduces the computation time. However, we need to be careful to only bin channels together which have frequency correlated $1/f$ noise, otherwise we will change the properties of the $1/f$ noise in the new averaged channel from the properties used as simulation input. In \autoref{fig:psRatio} we investigate the effect of frequency binning the HRD data on the ability of the noise-weighted map-maker to reduce the ratio between $1/f$ and white noise. It can be seen that at large $k$ values the choice of frequency binning has no effect on the accuracy of the map-maker as this is the regime where the white noise dominates and white noise can be reduced through simple averaging. At small $k$ values, where the $1/f$ noise power dominates, increasing the binning number from 2 to 5 (0.4\,MHz to 1.0\,MHz frequency resolution) shows a worthwhile improvement in the performance of the map-maker. Increasing the binning up to 10 channels (2\,MHz frequency resolution), however, largely amplifies the problem of $1/f$ noise implying that ten frequency channels span too large a frequency range for the $1/f$ noise properties to be consistent and so the time-correlated noise no longer averages down. Therefore we selected 1.0\,MHz as the preferred frequency resolution.

  \begin{figure}
    \centering
    \includegraphics[scale=0.56]{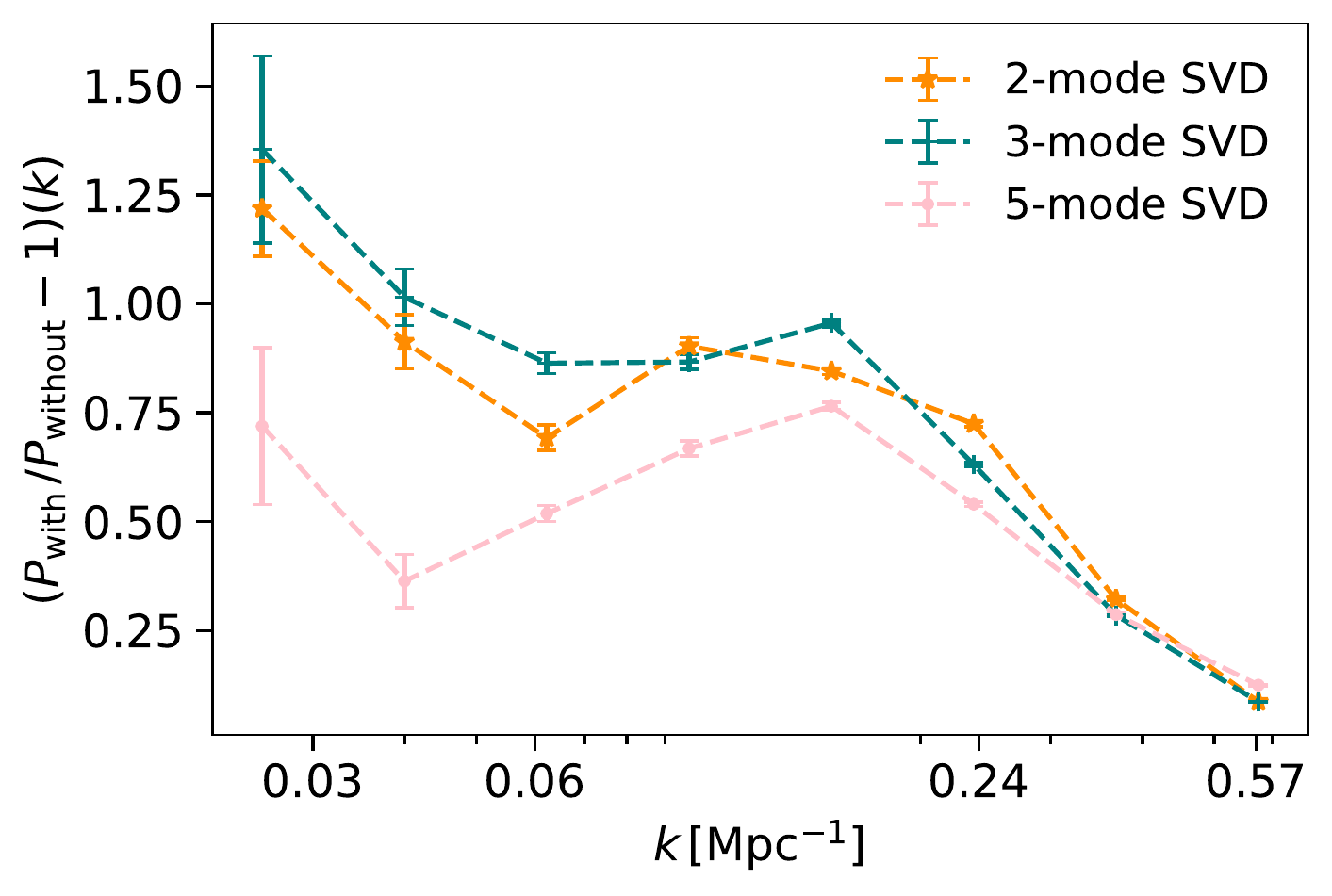}
    \caption{Auto-correlation spherically-averaged power spectra for the HRD scan strategy, noise-weighted maps. The maps were SVD cleaned and the fractional change in power is plotted for data with $1/f$ noise (i.e. $1/f$ noise plus white noise plus foregrounds plus the \hone\, signal) and for data without (i.e. white noise plus foregrounds plus the \hone\, signal).}
    \label{fig:ps}
 \end{figure}

\autoref{fig:ps} shows the effect of SVD cleaning on the auto-correlation power spectra. At this point we switch to using simulations which contain Galactic and extragalactic foregrounds from the GSM as these contributions are also correlated over frequency and so they will, alongside the $1/f$ noise, affect the impact of the SVD clean. We show the fractional difference between maps made from data both with and without the presence of $1/f$ noise. The HRD scan strategy data have been used and converted to maps using the noise-weighted technique. The larger the number of SVD modes removed, the closer the fractional power difference drops towards zero; meaning the smaller the impact of the $1/f$ noise above the residual foreground and white noise level. However SVD cleaning has also been shown to subtract the actual \hone \, signal itself. To help assess the level of over-cleaning/ signal-cleaning seen at low $k$, we enlist the help of cross-correlation power spectra.

    \begin{figure}
    \centering
    \includegraphics[scale=0.58]{ 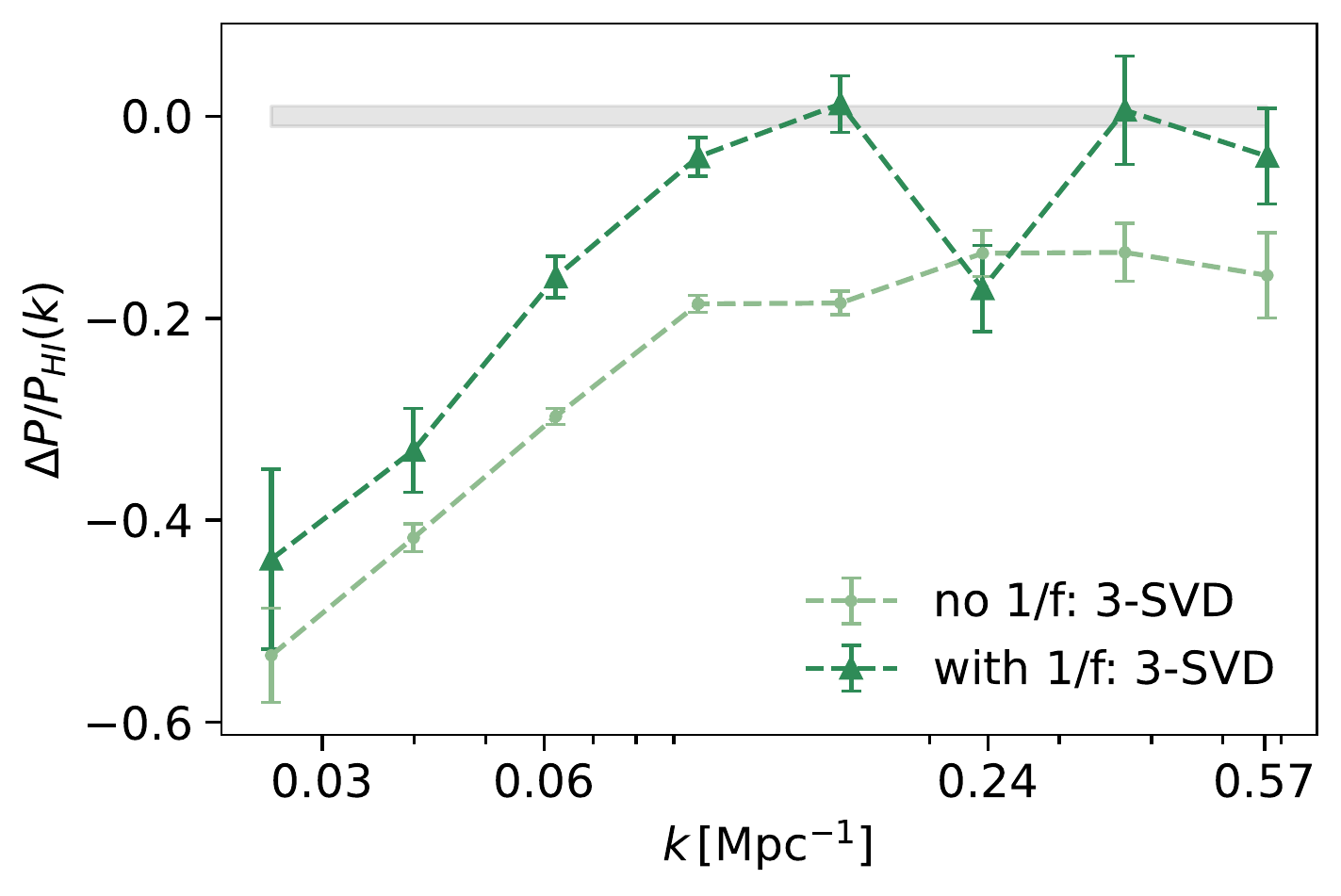} \\ 
    \includegraphics[scale=0.58]{ 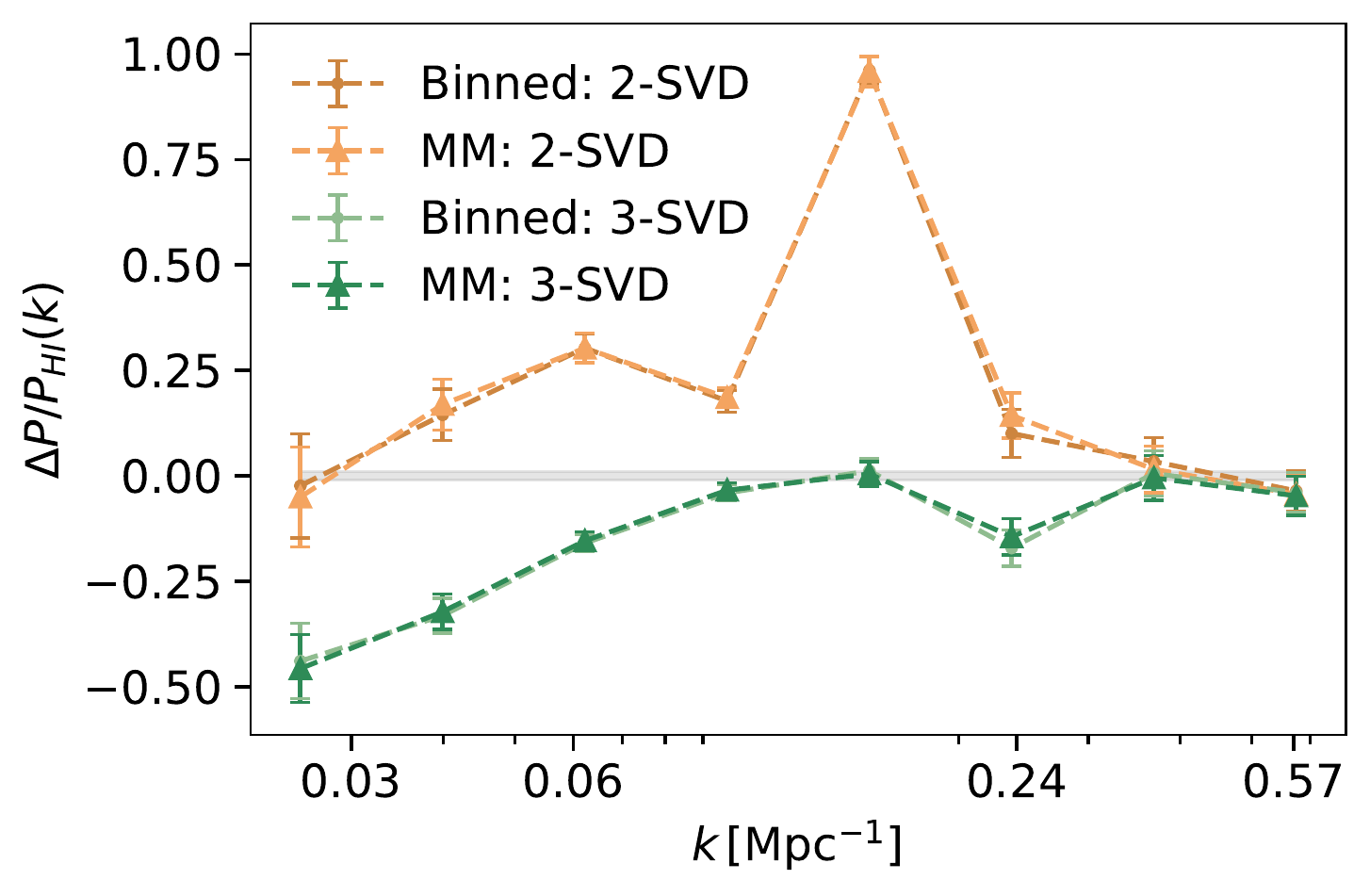} \\ 
     \includegraphics[scale=0.58]{ 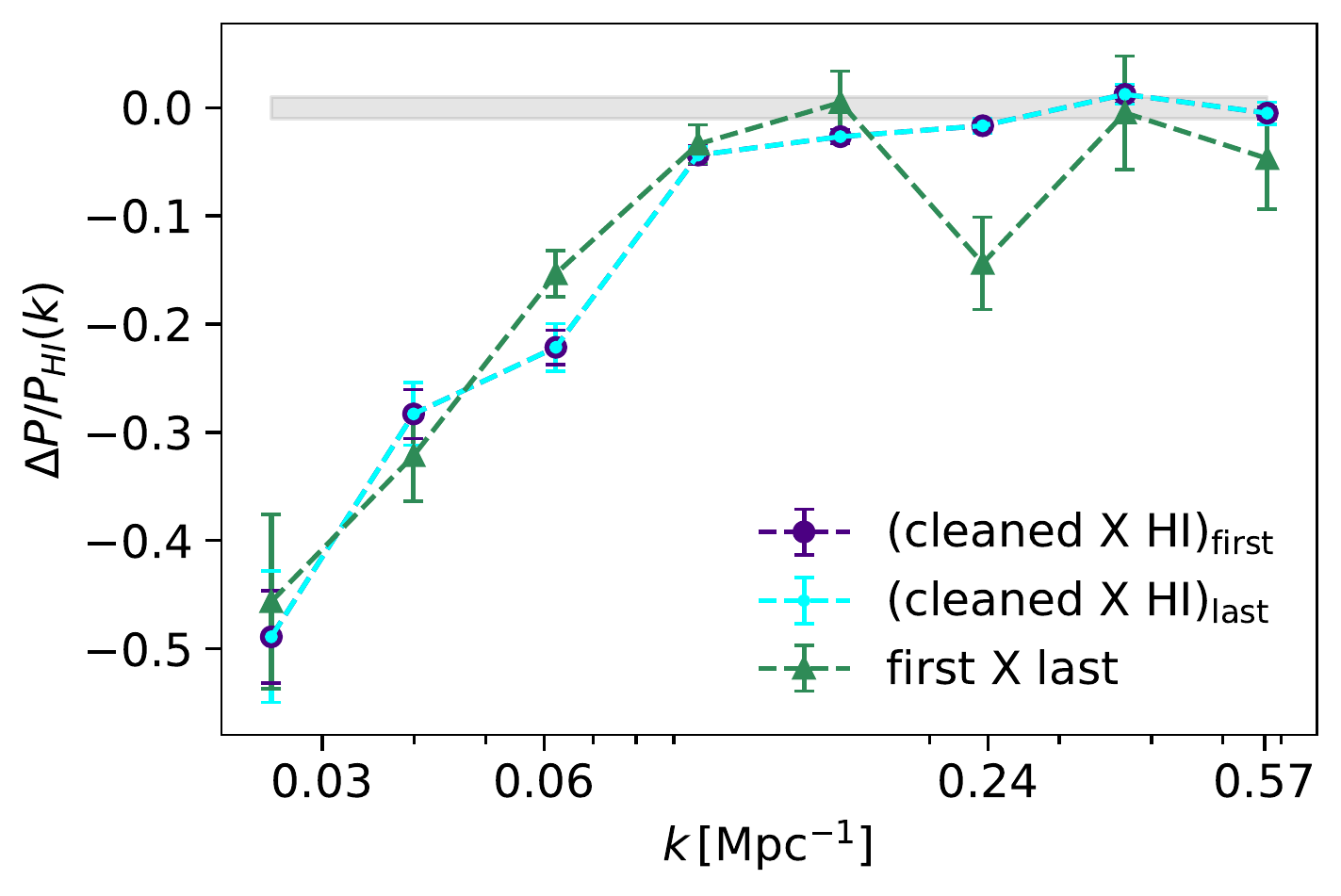} \\ 
    \caption{Fractional difference between the total and pure \hone \, cross-correlation power spectra for the HRD simulation data: binned maps with and without $1/f$ noise ({\it{top panel}}) and both binned and weighted maps containing $1/f$ noise ({\it{middle panel}}). The first eight observation blocks were cross-correlated with the last eight. The TOD for the results shown here all contain \hone \, emission, foregrounds and white noise. For the bottom panel the fractional power difference for the weighted, 3-mode SVD cleaned map containing $1/f$ noise is replotted against the fractional power difference for the cross-correlation between the first/last eight observation blocks (3-mode SVD cleaned and weighted map-making) and the \hone \, signal itself.}
    \label{fig:xps}
 \end{figure}

\autoref{fig:xps} shows the effect of SVD cleaning on the cross-correlation power spectra. The cross-correlation approach is probably the most optimal one for a \hone \, power spectrum detection since it will help to decorrelate systematics. Importantly, it will remove any noise bias, including the $1/f$ noise since it is independent between blocks. For the power spectra shown in \autoref{fig:xps}, the first eight HRD observation blocks were made into maps and then cross-correlated with the last eight observation blocks made into maps. The power difference ratio between the various power spectra and the pure \hone \, cross-correlation power spectrum is shown:
\begin{equation}
\frac{\Delta P(k)} {P_{\hone}(k)} = \frac{{\rm{cross}}(\hone \, + {\rm{white \, noise}} + 1/f + {\rm{foregrounds}})}{{\rm{auto}}(\hone)} - 1,
\end{equation}
where ${\rm{auto}}$ and ${\rm{cross}}$ signify the auto- and cross-correlation power spectrum. The power difference ratio value of zero is overlaid onto \autoref{fig:xps} using a solid gray line to highlight the optimum \hone \, power spectrum recovery. In this section the SVD cleaning was performed in the map-domain to represent the post-processing typically applied to intensity mapping data cubes. The three dimensional map cube containing the measurements from all dishes and all observation blocks was reduced to a two dimensional array of angular and frequency information in order for the frequency-frequency correlation matrix to be calculated. The Eigen-decomposition was then performed on this correlation matrix.

The top panel of {\autoref{fig:xps}} shows the cross-correlation spectra from maps made using simple binning of the TOD. Both power spectra were made from data containing \hone \, plus white noise plus foregrounds; the difference between the two spectra is the inclusion of $1/f$ noise. Both power spectra reveal that a 3-mode SVD clean would remove a significant amount of \hone \, signal as well as the frequency-correlated contaminants, thus biasing the measurement. However, such a bias could be compensated for by a technique such as the transfer function (see \autoref{sec:conc} for a discussion on the transfer function). The addition of $1/f$ noise both increases the error bars on the measurement as well as mixing with the residual foregrounds, which are common to both sets of observation blocks, to raise the power at all scales. This, in-turn, results in a lesser degree of over-cleaning. The middle panel of {\autoref{fig:xps}} reveals that at least 3 Eigen-modes were required to be removed in order to reach the \hone \, power spectrum level. Any fewer than three and there was residual power left over from the foreground emissions at medium scales. In the bottom panel of {\autoref{fig:xps}} we show the fractional power difference for the cross-correlation between the first/last eight observation blocks and the pure \hone \, signal itself. For this case the 3-mode SVD cleaned, weighted maps (contaminated with $1/f$ noise, foregrounds and white noise) were chosen to perform the cross-correlation. The fractional power difference for the cross-correlation between the cleaned maps and the \hone \, signal are comparable to the fractional power difference for the cross-correlation between the first and last eight observation maps; which we plot again in the bottom panel to illustrate this point. The cross-correlations with the pure \hone \, signal, however, can be seen to result in fractional difference power spectra with smaller errors bars as only one of the data sets involved in the cross-correlation for these spectra contain contaminants (residual $1/f$ and foreground emissions remaining after the SVD clean).

Regardless of the method of map-making used we see around a 40 per cent signal loss at large scales for a 3-mode SVD clean. This is consistent with the simulations analysis performed by \citet{bsazy}, where it was found that for $1/f$ noise which is perfectly correlated across frequency channels an SVD clean of 2-modes or more could remove all the excess power with a signal loss of only around 5 per cent. However if there is spectral structure to the \hone \, contaminants, for instance $1/f$ noise which is not perfectly correlated ($\beta > 0$), then the SVD clean struggles to remove the contaminants without a large loss in the measured \hone \, signal. \citet{bsazy} simulate perfectly correlated $1/f$ noise but then add in a non-smooth instrumental band-pass and find a \hone \, signal loss of up to 66 per cent, depending on the magnitude of the band-pass perturbation.

\section{Conclusions}
 \label{sec:conc}
 
This work has used simulated data to investigate the impact that scan strategy, SVD cleaning and noise-weighted map-making can have on the measurement of the \hone \, power spectrum from TOD containing $1/f$ noise. The simulations used reflected the parameters of the MeerKLASS \hone \, intensity mapping survey in terms of receiver temperature, spectral resolution and integration time. The magnitude and frequency correlation of the simulated $1/f$ noise used in the simulations was a realistic representation of the level of $1/f$ noise expected for the MeerKLASS data, as it was fitted from empirical 2019 pilot survey data. The discussion of whether or not to include noise diode firing events to calibrate out our $1/f$ noise is left for future work as the MeerKLASS calibration approach is still being honed.

MeerKLASS pilot data has so far been obtained using the HRD scan strategy, which scans the sky at 5 arcmin per second. This strategy was initially selected as it involves keeping the telescope at a constant scanning elevation; a useful tactic for ensuring minimum variation of the level of ground emission pick-up. Another common strategy is simply to integrate over specific telescope pointings for a longer time, this would entail moving the telescope in both azimuth and elevation. Through the use of these simulations we have shown that the HRD strategy is optimal for covering a large area of sky, without losing time samples to telescope slewing time, within a set observation time and restricting the effect of $1/f$ noise to the lowest $k$ values on the spherically averaged power spectra. This conclusion was also clear from \autoref{fig:ps2d} which only considered the additional power that the raw $1/f$ noise provided. In these 2D plots the random strategy was seen to spread the $1/f$ noise throughout the whole of $k$-space, while the HRD strategy contained it to large k-modes which are the modes most likely to be contaminated by Galactic foregrounds anyway. Future simulations will need to include contaminants other than the $1/f$ noise and foregrounds, such as resdiual radio frequency interference for example. In this work we considered the change in system temperature caused by changing observational elevation, which the random scan strategy does; however, we assumed a perfect modelling and removal of this elevation-dependant contribution. 

The TOD can be averaged into maps using either a simple binning or through finding a maximum-likelihood solution using the full noise covariance matrix. We implement both these techniques to assess whether the full noise analysis is a worthwhile strategy to help mitigate $1/f$ noise. While our $1/f$ noise is partially correlated across frequency, we anticipate that the empirical data processing pipeline will involve some degree of foreground cleaning aimed at removing frequency-correlated signal. Therefore we limit our map-making considerations to a per-frequency basis e.g. we use the full noise covariance matrix at a single frequency to make the 2D map at that frequency and then repeat this process for each frequency channel. Each of the 16 observation blocks, combined to cover the full RA and declination range in the map domain, have independent systematic noise properties to each other. Therefore, noise-weighted map-making for each individual frequency channel is only beneficial when two or more point samples from a single observation block are combined into a single map pixel. Fortunately it is not particularly computationally expensive for us to implement the full maximum likelihood solution, and in doing so we see a reduction in the contamination of the \hone \, power spectra by $1/f$ noise. This reduction is most noticeable, at around 30 per cent, for the raw levels of $1/f$ noise. 

As the MeerKLASS $1/f$ noise is partially correlated across frequency, component separation techniques which exploit frequency-frequency correlations can be employed to help remove some of the additional power added by this systematic. We included Galactic and extragalactic foregrounds, courtesy of the GSM, to assess the level of SVD-cleaning required. However, it should be noted that, after residual ground pick-up, residual RFI and calibration errors are included we are very likely to require a higher number of SVD modes to be removed from the data \citep{steve}, hence the $1/f$ noise should no longer be a dominant systematic. 

The main problem of SVD cleaning frequency-correlated contaminants is that, at the largest scales, part of the \hone \, signal is also removed. This over-cleaning can be seen in the cross-correlation power spectra presented in this paper. One solution for empirical data which has been foreground cleaned is to use the transfer function technique \citep{masui, tfunc2, laura, steveTF}, which aims to quantify the level of over-cleaning through simulations and then correct for it. $1/f$ noise, however, is unique as a frequency-correlated contaminate as it can actually be calibrated out of the TOD. Having focused in this work on finding the optimum scan and map-making strategies for MeerKLASS, future work will target the mitigation of $1/f$ noise through data calibration.    

\section*{Acknowledgements}

We acknowledge the use of the Ilifu cloud computing facility, through the Inter-University Institute for Data Intensive Astronomy (IDIA). The MeerKAT telescope is operated by the South African Radio Astronomy Observatory, which is a facility of the National Research Foundation, an agency of the Department of Science and Innovation. We acknowledge support from the South African Radio Astronomy Observatory and National Research Foundation (Grant No. 84156). SC is supported by a UK Research and Innovation Future Leaders Fellowship grant [MR/V026437/1]. 
This result is part of a project that has received funding from the European Research Council (ERC) under the European Union's Horizon 2020 research and innovation programme (Grant agreement No. 948764; PB). PB acknowledges support from STFC Grant ST/T000341/1.

\section*{Data Availability}
The analysis pipeline software is available on request. Data products are also available on request but unrestricted public access is subject to a 12-month proprietary period starting from the date of this publication. Access to the raw data used in the analysis is public (for access information please contact archive@ska.ac.za).

\balance 
\bibliographystyle{mnras}
\bibliography{refs} 



\appendix
\section{Fitting 1\,/\,\lowercase{\textit{f}} noise from observational data}
\label{appendix:apA}

    \begin{figure}
    \centering
    \includegraphics[scale=0.55]{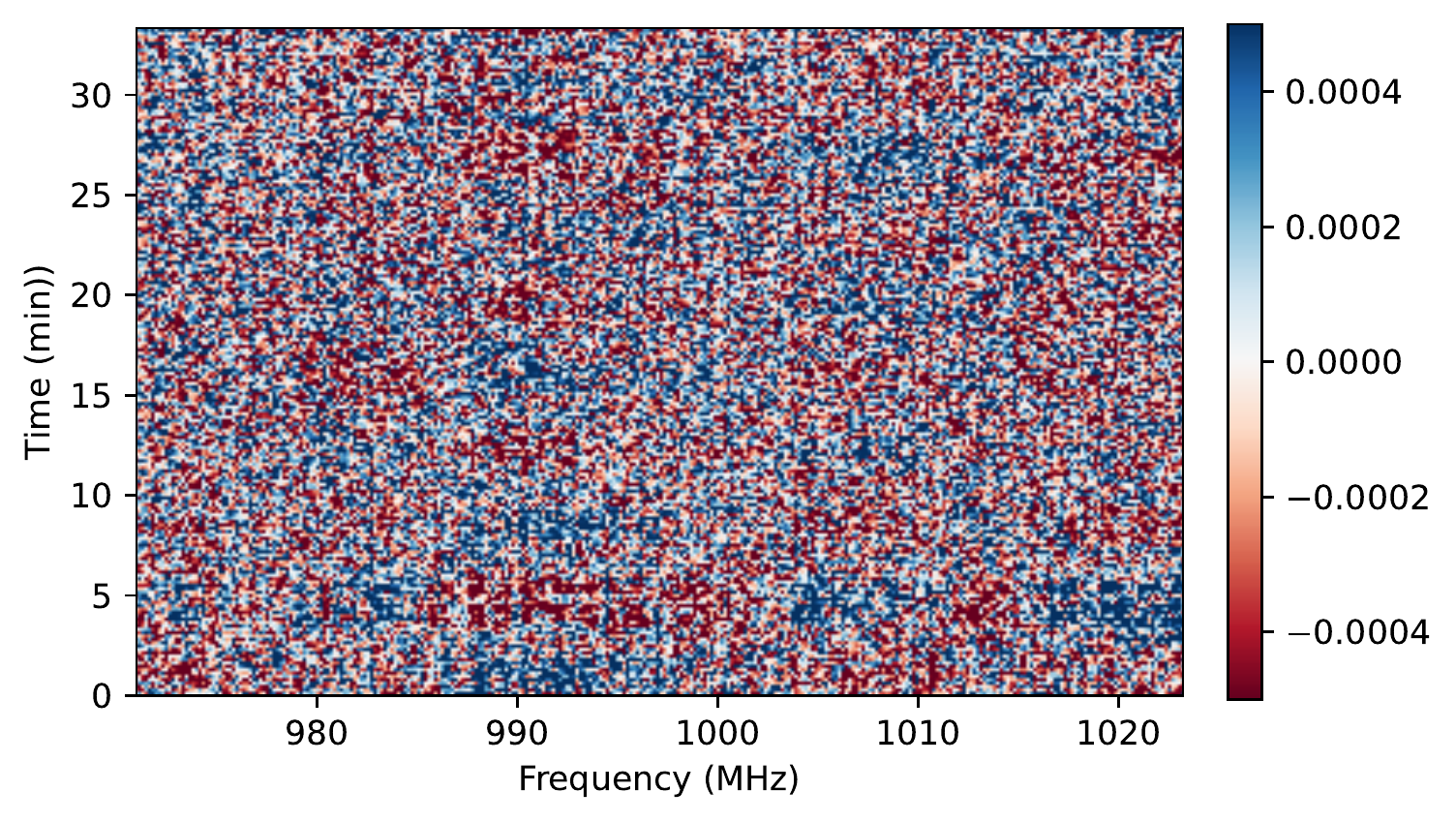}\\
    \includegraphics[scale=0.55]{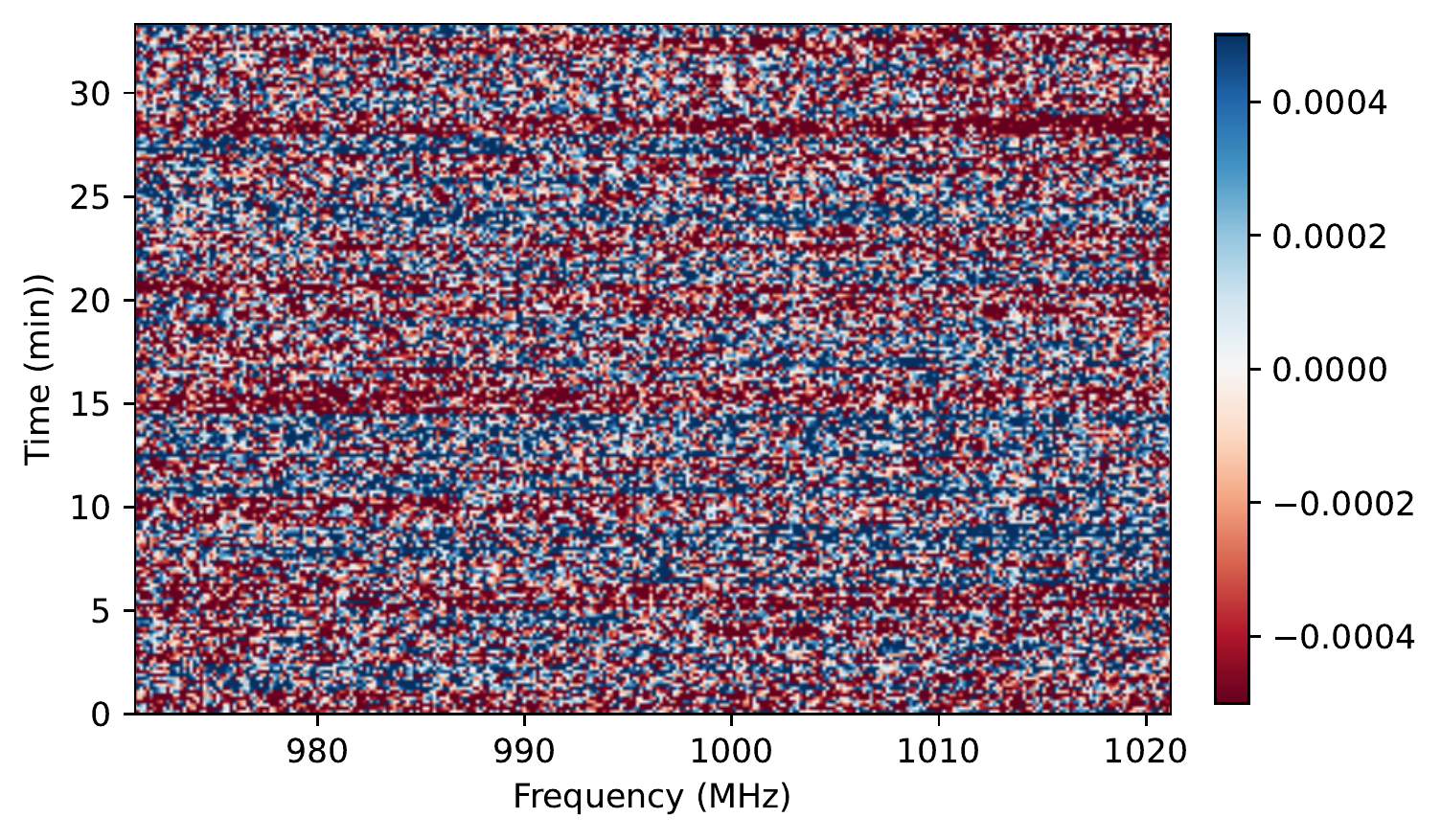}\\
    \includegraphics[scale=0.55]{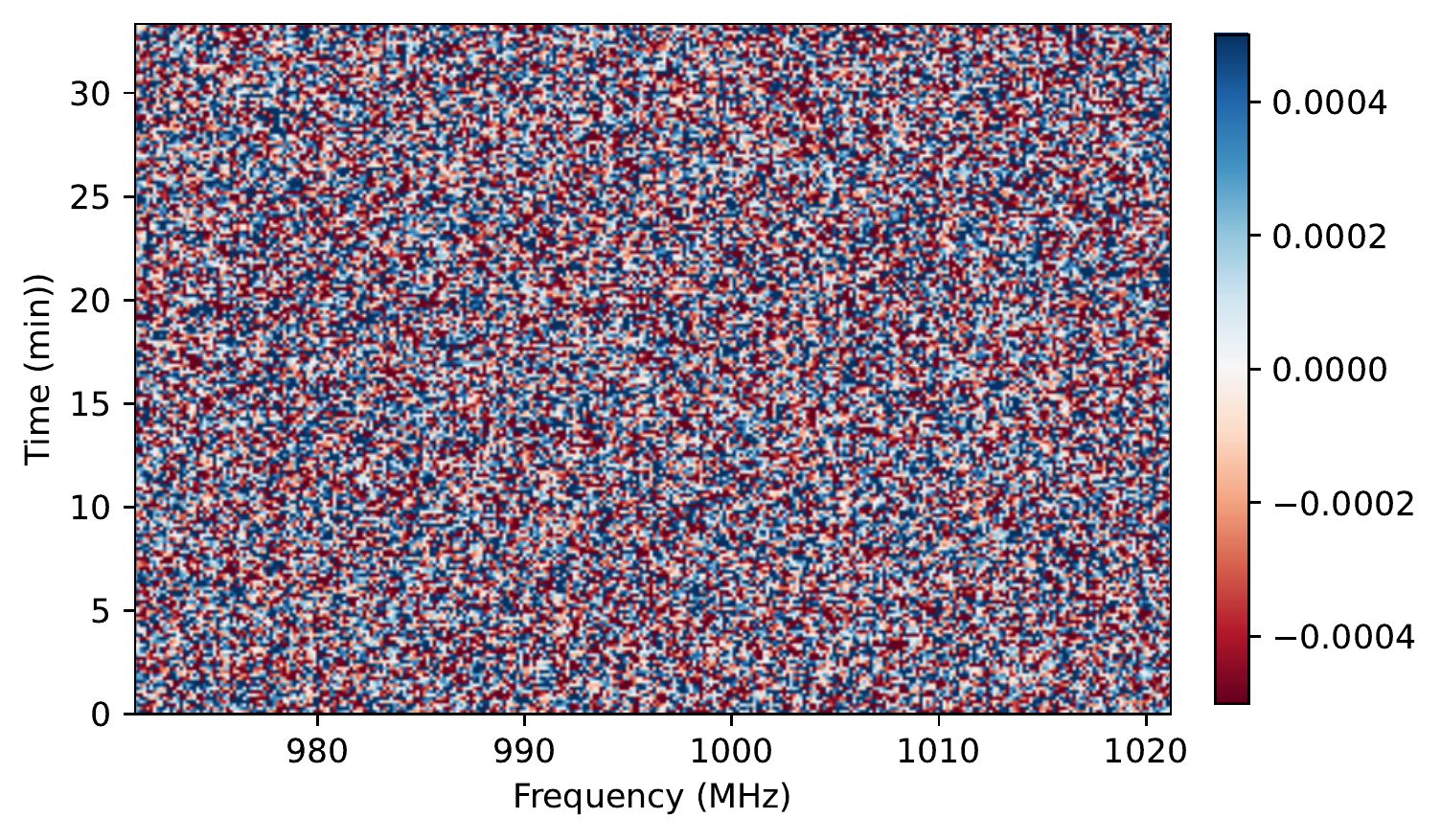}\\
    \caption{{\it{Top:}} Normalised (subtraction of and division by the mean), un-calibrated data from the MeerKAT pilot survey; receiver m002. The sky signal has been isolated using the cross-correlation between receivers m000 and m0002 and removed from the m0002 data. {\it{Middle:}} An example waterfall plot of the normalised simulated $1/f$ plus white noise data for a single receiver. {\it{Bottom:}} An example waterfall plot of the normalised simulated white noise data for a single receiver.}
    \label{fig:cleaned}
 \end{figure}

The $1/f$ noise simulator simply requires $\alpha$, $\beta$ and $f_{k}$ values to produce noise TOD. To ensure the realism of our simulations, these parameter values were taken from the MeerKLASS 2019 pilot survey data; specifically the telescope scanning part of the 90 minute observation conducted on 25$^{{\rm{th}}}$ February 2019 \citep{wang21}. As the empirical data contain not only the \hone \, signal, the white and $1/f$ noise components but also radio frequency interference (RFI), diffuse Galactic foregrounds and extragalactic point sources, the data require a pre-processing step before the $1/f$ noise can be fit. 

\subsection{Processing the observational data}
To isolate the $1/f$ signal from observational data we employed the method detailed in \citet{hu21}. This method takes advantage of the fact that when there are two (or more) receiver measurements of the sky signal, both receivers measure an identical sky signal but differing noise (since they are statistically independent).
After cross-correlating these two receivers, an SVD clean will isolate the sky signal which can then be subtracted from the TOD of each receiver leaving only the white noise plus $1/f$ noise fluctuations in each timestream.

For the MeerKAT analysis, 12 receivers (both horizontally and vertically
polarised) were chosen to characterise typical 1/f parameter values. In place of using a radio frequency interference flagger, we simply picked out the `cleanest' dishes by a quick visual inspection of the time ordered data. We specifically used data which had not been calibrated because data multiplied by gains will have different $1/f$ noise properties to the raw data. The selected frequency range was 250 channels between 960.5 and 1012.8\,MHz and an RFI free patch of the TOD spanning 33.3 minutes was selected. As we wanted to quantify the fluctuations around the time-average, the data were normalised at each frequency by subtraction and division by the mean across time. Meaning that the normalised TOD at a single frequency channel, channel `$\nu_{1}$' for instance is calculated as: 
\begin{equation}
\label{eq:div}
{\bf{\delta Y}}(t, \nu_{1}) = \frac{{\bf{Y}}(t, \nu_{1}) - \overline{Y(t, \nu_{1})}}{\overline{Y(t, \nu_{1})}}.
\end{equation}
We expected that the division would remove the band-pass features so that the fluctuations seen in frequency were not due to the band-pass. This particular stint of pilot data were taken using noise diode injections therefore the noise diode events had to be flagged and replaced with zeros. An SVD was then applied to the cross-correlation of pairs of receivers (coined receivers `A' and `B'): 
\begin{equation}
{\bf{C_{v^{A}v^{B}}}} = \langle {\bf{Y_{A}}} {\bf{Y_{B}}}^{t} \rangle = {\bf{U_{A}}} {\bf{\Lambda}} {\bf{U_{B}}}^{t},
\end{equation}
where ${\bf{Y_{A/B}}}$ is the data matrix of TOD at each frequency. We select the ten brightest modes of the SVD to represent the sky signal; \citet{hu21} find that fewer than ten modes are sufficient to represent the sky signal. We use the data from receiver B, with the sky signal subtracted to represent the typical noise properties for MeerKLASS. \autoref{fig:cleaned} presents a waterfall plot of one of the cleaned receiver normalised data alongside both an example plot of the $1/f$ plus white noise and an example plot of white noise only from our simulations. The structure of $1/f$ noise appears as horizontal stripes running parallel to the frequency axis.  

\subsection{The 1\,/\,\textit{f} noise parametric fit}
\label{sec:thefit}

Once an estimate of the \hone \, signal plus white and $1/f$ noise had been obtained from the observational data, the $1/f$ parametric fit could be used to determine the values for $\alpha$, $\beta$ and $f_{k}$ we wished to use to generate simulated $1/f$ noise.

We chose to split the 2D power spectrum into its temporal and spectral components and fit to each separately (this is possible because the 2D power spectrum is separable in each direction). First we performed the fit to the time-correlated $1/f$ noise: 
\begin{equation}
S(f) = c_{t} \left( 1 + \left(\frac{f_{k}}{f}\right)^{\alpha} \right).
\end{equation}
The free parameters allowed for the fit were $\alpha$, $f_{k}$ and a normalisation constant $c_{t}$.
We binned the data across five channels in frequency, changing the resolution from 0.2\,MHz to 1\,MHz and the number of channels available from 250 to 50 in order to increase the signal-to-noise ratio for the fit. The power along time for each of these 50 channels was then calculated:
\begin{equation}
S(f) = \frac{\delta t}{N_{t}} \fft({\delta Y(t})) \fft({\delta Y(t}))^{*},
\end{equation}
where $N_{t}$ is the number of TOD samples.
These 50 power spectra were then used to calculate the mean power value and the standard error on this mean was used for the fit. Note that the scatter calculated across these 50 power spectra may not fully account for the error in the cases where the correlations between the different binned channels is strong. This should not however affect the best fit values.
This fit gives one measurement of $\alpha$ and $f_{k}$ from the observational data for each receiver. 

To fit the spectral component of the 2D power spectrum: 
\begin{equation}
S(\tau) = c_{s} \left( 1 + \left(\frac{\tau_{0}}{\tau}\right)^{\frac{1-\beta}{\beta}} \right),     
\end{equation}
the TOD in each frequency channel in the 0.2\,MHz resolution data were binned across 50 time dumps, giving a 100\,s time resolution (since each time dump is 2\,s). The free parameters for the spectral fit are $\beta$, $\tau_{0}$ and a normalisation constant $c_{s}$.
As before, the mean power value and the standard error on this mean was used for the fit.

  \begin{figure}
 \centering
   \includegraphics[scale=0.52]{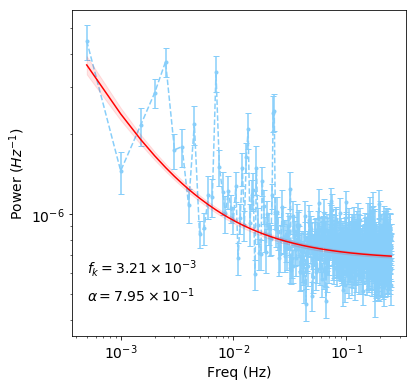}
   \includegraphics[scale=0.52]{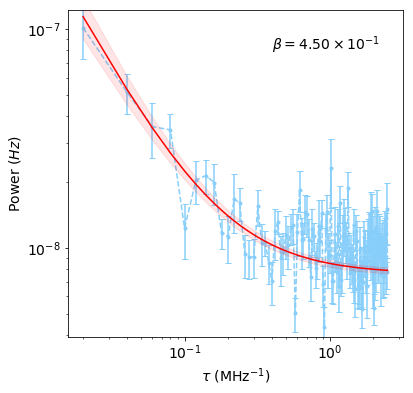}\\
    \caption{Parametric fit to the temporal ({\it{top}}) and spectral ({\it{bottom}}) power spectrum for one MeerKAT receiver. As the data have been normalised by subtraction of and division by the mean, the power units do not include Kelvin. Shaded regions show the marginalised 1$\sigma$ confidence level for the model.}
    \label{fig:fftfits}
 \end{figure}

An example of the parametric fit to both the temporal and spectral power spectra is given in \autoref{fig:fftfits}. The error on the model fit ($\Delta M$) is due to the combined errors on the fitted parameter triplets; either ($(\alpha, f_{k}, c_{t})$) for the temporal fit or ($(\beta, t_{0}, c_{s}$)) for the spectral fit. As the individual parameter errors are correlated the 3D Gaussian posterior probability distribution ($p(M_{0}, g)$) must be calculated for the full parameter space: 
\begin{align}
\langle M(x) \rangle &= \int \int {\rm{func}}(M_{0}, g; x) p(M_{0}, g) \, d M_{0} \,  dg, \\
\langle M^{2}(x) \rangle &= \int \int {\rm{func}}^{2}(M_{0}, g; x) p(M_{0}, g) \, d M_{0} \, dg, \\
\Delta M(x) &= \sqrt{\langle M^{2}(x)\rangle - \langle M(x)\rangle^{2}}, 
\end{align}
where for the temporal power spectrum $g = (\alpha, f_{k}, c_{t})$ and $x = f$ and for the spectral power spectrum $g = (\beta, t_{0}, c_{s}$) and $x = \tau$. The posterior distribution has been normalised to unity.

 \begin{figure}
 \centering
   \includegraphics[scale=0.55]{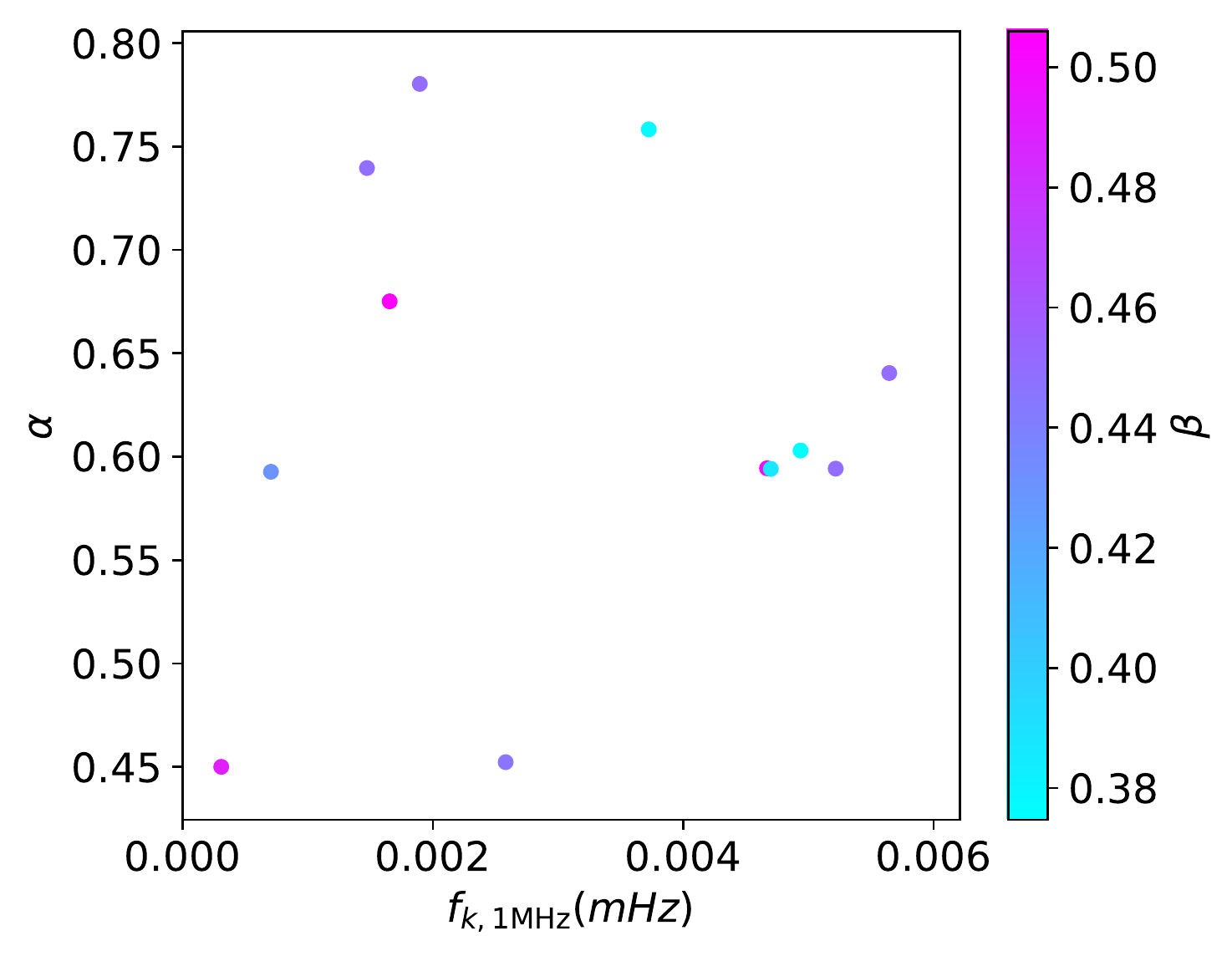}\\
    \caption{Fitted $1/f$ noise parameters for twelve of the MeerKAT receivers.}
    \label{fig:therecs}
 \end{figure}

Having fit an estimate for $\alpha$, $\beta$ and $f_{k}$ from each receiver, we can then use these parameters within our simulations. \autoref{fig:therecs} displays the parameter triplets ($\alpha$, $\beta$ and $f_{k}$) fitted for twelve receivers. For computational ease we do not simulate all 120 receivers, instead we simulate 12 receivers and divide the white and $1/f$ noise TOD by $\sqrt{10}$ to enable our simulated receivers to represent ten receivers each. We can see that there is reasonable spread of parameters values; this level of spread was also shown in the MeerKAT receiver noise analysis of \cite{li21}. 

The $f_{k}$ parameter has been measured from TOD with a frequency resolution of 1\,MHz. The derivation of the relationship between the knee frequency at resolutions $\delta \nu$ and $\delta \nu^{\prime}$: 
\begin{equation}
\log{f_{k}} = \log{f_{k^{\prime}}} + \frac{1}{\alpha} \log(\frac{K \delta \nu}{K^{\prime} \delta \nu^{\prime}})  
\end{equation}
is given in Appendix A of \citet{li21}.


\bsp	
\label{lastpage}
\end{document}